 \documentclass[11pt]{article}

\usepackage{amsmath,amssymb}

\newtheorem{teor}{Theorem}
\newtheorem{prop}{Proposition}
\newtheorem{corol}{Corollary}
\newtheorem{lem}{Lemma}
\newtheorem{definition}{Definition}

\def\beq{\begin{equation}}
\def\eeq{\end{equation}}
\def\bea{\begin{eqnarray}}
\def\eea{\end{eqnarray}}
\def\beann{\begin{eqnarray*}}
\def\eeann{\end{eqnarray*}}
\def\beasn{\begin{sneqnarray}}
\def\eeasn{\end{sneqnarray}}
\def\ben{\begin{enumerate}}
\def\een{\end{enumerate}}
\def\bit{\begin{itemize}}
\def\eit{\end{itemize}}
\def\proof{ ({\sl Proof\/}) }
\def\derpar#1#2{\displaystyle\frac{\partial{#1}}{\partial{#2}}}
\def\derpars#1#2#3{\displaystyle\frac{\partial^2{#1}}{\partial{#2}\partial{#3}}}

\def\qed{\ifvmode\Realemovelastskip\fi
{\unskip\nobreak\hfil\penalty50\hbox{}\nobreak\hfil \hbox{\vrule
height1.2ex width1.2ex}\parfillskip=0pt \finalhyphendemerits=0
\par\smallskip}}

\def\vf{\mathfrak{X}}
\def\df{{\mit\Omega}}

\def\d{{\rm d}}

\def\Real{\mathbb{R}}
\def\R{\mathbb{R}}

\def\Tan{{\rm T}}
\def\Lie{\mathop{\rm L}\nolimits}
\def\inn{\mathop{i}\nolimits}
\def\Cinfty{{\rm C}^\infty}
\def\tabaddress#1{{\small\it\begin{tabular}[t]{c}#1
\\[1.2ex]\end{tabular}}}
\def\qed{\ifvmode\removelastskip\fi
{\unskip\nobreak\hfil\penalty50\hbox{}\nobreak\hfil \hbox{\vrule
height1.2ex width1.2ex}\parfillskip=0pt \finalhyphendemerits=0
\par\smallskip}}

\newcommand{\ds}{\displaystyle}

\title{SYMMETRIES AND CONSERVATION LAWS IN THE G\"{U}NTHER $k$-SYMPLECTIC
FORMALISM OF FIELD THEORY}
\author{\sc Narciso Rom\'an-Roy\thanks{{\bf e}-{\it mail}:
  nrr@ma4.upc.edu}
  \\
  \tabaddress{Departamento de Matem\'atica Aplicada IV\\
  Edificio C-3, Campus Norte UPC\\
  C/ Jordi Girona 1. 08034 Barcelona. Spain}
  \\
{\sc Modesto Salgado\thanks{{\bf e}-{\it mail}:
modesto@zmat.usc.es},
     Silvia Vilari\~no\thanks{{\bf e}-{\it mail}:
svfernan@usc.es}} \\
 \tabaddress{Departamento de Xeometr\'{\i}a e Topolox\'{\i}a\\
 Facultade de Matem\'{a}ticas,
    Universidade de Santiago de Compostela,\\
    15706-Santiago de Compostela, Spain}}

\begin{document}

\maketitle

\pagestyle{myheadings}

\thispagestyle{empty}

\begin{abstract}
This paper is devoted to studying symmetries of $k$-symplectic
Hamiltonian and Lagrangian first-order classical field theories. In
particular, we define symmetries and Cartan symmetries and study
the problem of associating conservation laws to these symmetries, stating and
proving Noether's theorem  in different situations for the
Hamiltonian and Lagrangian cases. We also characterize
equivalent Lagrangians, which lead to an introduction of Lagrangian gauge
symmetries, as well as analyzing their relation with Cartan symmetries.
\end{abstract}


\bigskip

\vbox{\raggedleft AMS s.\,c.\,(2000): 70S05, 70S10, 53D05}\null

\markright{\sc N. Rom\'an-Roy {\it et al\/},
    \sl Symmetries and conservation laws in $k$-symplectic field theory}

  \clearpage

\tableofcontents


\section{Introduction}

 G\"{u}nther's paper  \cite{gun} gives a
geometric Hamiltonian formalism for field theories. The crucial
device is the introduction of a vector-valued generalization of a
symplectic form, called a polysymplectic form. One of the advantages
of this formalism is that one only needs the tangent and cotangent
bundle of a manifold to develop it. In \cite{fam} G\"{u}nther's
formalism has been revised and clarified. It has been shown that the
polysymplectic structures used by G\"{u}nther to develop his formalism
could  be replaced by the $k$-symplectic structures defined by Awane
\cite{aw1,aw3}. So this formalism is also called $k$-symplectic
formalism.

The $k$-symplectic formalism is the generalization to field
theories of the standard symplectic formalism in Mechanics, which
is the geometric framework for describing autonomous dynamical
systems. In this sense, the $k$-symplectic formalism is used to
give a geometric description of certain kinds of field theories: in
a local description, those theories whose Lagrangian does not depend on the
base coordinates, denoted by $(t^1,\ldots,t^k)$ (in many of these, the space-time
coordinates); that is, the $k$-symplectic formalism is only valid for Lagrangians
$L(q^i,v^i_A)$ and  Hamiltonians $H(q^i,p^A_i)$ that depend on
the field coordinates $q^i$ and  on the partial derivatives of the
field $v^i_A$, or the corresponding moment $p_i^A$. A natural extension of this formalism is the
so-called $k$-cosymplectic formalism, which is the generalization
to field theories of the cosymplectic formalism
geometrically describing non-autonomous mechanical systems (this description
can be found in \cite{mod1,mod2}). This formalism is devoted to describing field
theories involving the coordinates $(t^1,\ldots ,t^k)$
on the Lagrangian $L(t^A,q^i,v^i_A)$ and on the Hamiltonian $H(t^A,q^i,p^A_i)$.

Let us remark here that the polysymplectic formalism developed by
Sardanashvily \cite{Sarda2}, based on a vector-valued form defined on
some associated fiber  bundle, is a different description of
classical field theories of first order than the polysymplectic
(or $k$-symplectic) formalism proposed by G\"{u}nther
(see also \cite{Kana} for more details).
 We must also remark that the soldering form on the linear
frames bundle is a polysymplectic form, and its study and
applications to field theory, constitute the $n$-symplectic geometry
developed by L. K. Norris in \cite{McN,No2,No3,No4,No5}.

An alternative way to derive the field equations is to use the
so-called multisymplectic formalism, developed by  Tulczyjew's
school in Warsaw (see \cite{Kijo,KijoSz,KijTul,Snia}), and
independently by Garc\'{\i}a and P\'{e}rez-Rend\'{o}n \cite{GP1,GP2} and
Goldschmidt and Sternberg \cite{GS}. This approach was revised by
Martin \cite{Mart1,Mart2} and Gotay et al \cite{Go1,Go2,Go3,Gymmsy}
and more recently by Cantrijn {\it et al} \cite{Cant1,Cant2}.

The  aim of this paper is  to study  symmetries and conservation
laws on first-order classical field theories, both for the Lagrangian and Hamiltonian formalisms,
 using G\"{u}nther's $k$-symplectic description,
and considering only the regular case. The study of symmetries of
$k$-symplectic Hamiltonian systems, is, of course, a topic of great
interest. The general pro\-blem of a group of symmetries acting on a
$k$-symplectic manifold and the subsequent theory of reduction
 has recently been analyzed in \cite{fam}.
Here, we recover the idea of {\sl conservation law}
 or {\sl conserved quantity}, and state
 Noether's theorem for Hamiltonian and Lagrangian systems in $k$-symplectic field theories.
 Thus, a large part of our discussion is a generalization of
 the results obtained for non-autonomous mechanical systems
 (see, in particular, \cite{LM-96}, and references quoted therein).
We further remark that the problem of symmetries in field theory
has also been analyzed using other geometric frameworks,
such as the {\sl multisymplectic models}
(see, for instance, \cite{EMR-99,Gymmsy,LMS-2004}.

The organization of the paper is as follows: Sections 2 and 3 are
devoted to the study of symmetries and conservation laws in
Hamiltonian $k$-symplectic field theory and Lagrangian
$k$-symplectic field theory, respectively. In particular, in
Sections 2.1 and 2.2 we develop the Hamiltonian formalism. So, in
Section 2.1 the field theoretic phase space of moments is
introduced as the Whitney sum $(T^1_k)^*Q$ of $k$-copies of the
cotangent bundle $T^*Q$ of a manifold $Q$. This space is the
canonical example of polysymplectic manifold introduced by G\"{u}nther
and $k$-symplectic manifolds introduced by Awane \cite{aw1,aw2,aw3}.
 In Section 2.2, the Hamiltonian $k$-symplectic formalism is described.
In Section 2.3 we obtain the main results of this Section:
after introducing different kinds of symmetries and their relation,
we can associate to some of them (the so-called {\sl Cartan symmetries}\/)
a conservation law ({\sl Noether's Theorem\/}).

Concerning the Lagrangian formalism (Section 3), the field
theoretic state space of velocities is introduced in Section 3.1
as the Whitney sum $T^1_kQ$ of $k$-copies of the tangent bundle $TQ$
of a manifold $Q$. This manifold has a canonical $k$-tangent
structure defined by  $k$ tensors fields of type $(1,1)$. The
$k$-tangent manifolds were introduced in de Le\'{o}n {\it et al.}
\cite{mt1,mt2}, and they generalize the tangent manifolds. A
geometric interpretation of the second order partial differential
equations is also given. Here we show that these equations can be
characterized using the canonical $k$-tangent structure of $T^1_kQ$,
which generalizes the case of Classical Mechanics. The Lagrangian
formalism  is developed  in Section 3.2, where the canonical
$k$-tangent structure of $T^1_kQ$ is used for its construction
instead of the Legendre transformation as in G\"{u}nther \cite{gun}. In
Section 3.3  we discuss symmetries and conservation laws in the
Lagrangian case, obtaining results analogous to those in Section
2.3, including the corresponding Noether's theorem.
 Finally, in Sections 3.4 and
3.5 we introduce the notion of gauge equivalent Lagrangians,
showing that they give the same solutions to the Euler-Lagrange
equations. This leads to the introduction of the so-called {\sl Lagrangian
gauge symmetries}, and to stating a version of the Noether theorem for
a particular class of them.

All manifolds are real, paracompact, connected and $C^\infty$. All
 maps are $C^\infty$. Sum over crossed repeated indices is understood.

\section{Hamiltonian $k$-symplectic case}
\protect\label{Hksc}

\subsection{Geometric elements}
\protect\label{21}

\subsubsection{The cotangent bundle of $k^1$-covelocities of a manifold.
Canonical  structures}

 Let $Q$ be a differentiable manifold of dimension  $n$
and $\tau^*_Q: T^*Q \to Q$ its cotangent bundle . We denote by
$(T^1_k)^*Q= T^*Q\oplus \stackrel{k}{\dots} \oplus T^*Q$ the Whitney
sum of $k$ copies of $T^*Q$, with projection map $\tau^*\colon
(T^1_k)^* Q \to Q$, $\tau^* (\alpha_{1_q},\ldots ,\alpha_{k_q})=q$.

The manifold  $(T^1_k)^*Q$ can be canonically
identified with the vector bundle $J^1(Q,\Real^k)_0$ of
$k^1$-covelocities of the manifold $Q$, the manifold of $1$-jets
of maps $\sigma\colon Q\to\R^k$ with target at $0\in \Real^k$ and projection map
$\tau^*\colon J^1(Q,\Real^k)_0\to Q$, $\tau^* (j^1_{q,0}\sigma)=q$; that is,
\[
\begin{array}{ccc}
J^1(Q,\Real^k)_0 & \equiv & T^*Q \oplus \stackrel{k}{\dots} \oplus T^*Q \\
j^1_{q,0}\sigma & \equiv & (d\sigma^1(q), \dots ,d\sigma^k(q))
\end{array}
\]
where $\sigma^A= \pi_A \circ \sigma:Q \longrightarrow \Real$ is
the $A$-th component of $\sigma$, and  $\pi_A\colon \Real^k \to \Real$
are the canonical projections, $1\leq A \leq k$.
For this reason,  $(T^1_k)^*Q$ is also called
 {\sl the bundle of $k^1$ covelocities of the manifold $Q$.}

If $(q^i)$ are local coordinates on $U \subseteq Q$,  then the
induced local coordinates $(q^i , p_i)$, $1\leq i \leq n$, on
$T^*U=(\tau_Q^*)^{-1}(U)$,   are given by
$$
q^i(\alpha_q)=q^i(q), \quad p_i(\alpha_q)=
\alpha_q\left(\frac{\partial}{\partial q^i}\Big\vert_q\right)
$$
and the induced  local coordinates $(q^i , p^A_i),\, 1\leq i \leq
n,\, 1\leq A \leq k$, on $(T^1_k)^*U=(\tau^*)^{-1}(U)$ are
$$
  q^i(\alpha_{1_q},\ldots , \alpha_{k_q})=q^i(q),\qquad
p^A_i(\alpha_{1_q},\ldots , \alpha_{k_q})=
\alpha_{A_q}\left(\frac{\partial}{\partial q^i}\Big\vert_q\right)\, .
$$

If $\tau^*_Q\colon T^*Q \to Q$ is the canonical projection,
the Liouville $1$-form $\theta\in\df^1(T^*Q)$ is given by
$$
\theta(\alpha_q)(\widetilde{X}_{\alpha_q})
=\alpha_q((\tau^*_Q)_*(\alpha_q)(\widetilde{X}_{\alpha_q})), \quad
\alpha_q\in T^*Q, \,\, \widetilde{X}_{\alpha_q}\in
T_{\alpha_q}(T^*Q),
$$
then $\omega=-\d\theta$ is the canonical symplectic structure in $T^*Q$,
and therefore we define
$$
\omega^A = (\tau^*_A)^*\omega, \quad 1\leq A \leq k\, ,
$$
where $\tau^*_A\colon (T^1_k)^*Q \rightarrow T^*Q $ is the
canonical projection on the $A^{th}$-copy  $T^*Q$ of $(T^1_k)^*Q$.
Of course, $\omega^A =-\d\theta^A$, where $\theta^A=(\tau_A^*)^*\theta$.
Thus, the {\sl canonical $k$-symplectic structure} on $(T^1_k)^*Q$ is
given by the family  $(\omega^A,V;1\leq A\leq k)$, in $(T^1_k)^*Q$,
where $V =\ker(\tau^*)_*$ (see \cite{aw1,aw3,fam}).

As the canonical symplectic structure on $T^*Q$ is locally given by
$\omega=-\d (p_i \, \d q^i )= \d q^i\wedge\d p_{i}$,
then the canonical forms $\omega^A$ in $(T^1_k)^*Q$ are locally given by
\begin{equation}
\label{canosym}
\omega^A= -\d\theta^A=- \d(p^A_i \,\d q^i )=\d q^i\wedge\d p^A_{i}  \, .
\end{equation}

It is interesting to recall that the {\sl canonical polysymplectic structure} in  $(T^1_k)^*Q$
introduced by  G\"{u}nther \cite{gun} is
the closed non-degenerate $\Real^k$-valued $2$-form
$\bar{\omega}=\displaystyle  \omega^A \otimes r_A$,
 where $\{r_1, \ldots, r_k\}$ denotes the canonical basis of $\Real^k$.

\subsubsection{Complete lift of diffeomorphisms and vector fields from $Q$ to $(T^1_k)^*Q$}

Now, let $\varphi\colon Q \to Q$ be a diffeomorphism, then the cotangent map
$T^*\varphi\colon T^*Q\to T^*Q$ is given by
$T^*\varphi(\alpha_q)=\alpha_q\circ \varphi_*(\varphi^{-1}(q))$.
We define the {\sl canonical prolongation} of $\varphi$ to $(T^1_k)^*Q$ as the map
$(T^1_k)^*\varphi:(T^1_k)^*Q \to (T^1_k)^*Q$ given by
$$
(T^1_k)^*\varphi({\alpha_1}_q,\ldots,{\alpha_k}_q)=
(T^*\varphi({\alpha_1}_q),\ldots,T^* \varphi({\alpha_k}_q)) \quad ,\quad
\mbox{\rm for $({\alpha_1}_q,\ldots,{\alpha_k}_q)\in (T^1_k)^*Q$, $q\in Q$} \, .
$$
If $Z$ is a vector field on $Q$, with local $1$-parametric group of
transformations $h_s\colon Q \to Q$ then the local $1$-parametric
group of transformations $(T^1_k)^*(h_s)\colon (T^1_k)^*Q
\to(T^1_k)^*Q$ generates a vector field $Z^{C*}$ on $(T^1_k)^*Q$,
which is called the {\sl canonical lift} of $Z$ to $(T^1_k)^*Q$. If
$Z=Z^i\ds\frac{\partial}{\partial q^i}$, the local expression of $
Z^{C*}$ is
$$
Z^{C*}=Z^i\frac{\partial}{\partial q^i} \, - \, p_j^A \ds
\frac{\partial Z^j} {\partial q^k} \frac{\partial}{\partial p_k^A}
\; .
$$

The canonical liftings or prolongations of diffeomorphisms and
vector fields on the base manifold $Q$ to $(T^1_k)^*Q$ have the
following properties:

\begin{lem}
\label{fiprop1h}
\ben
\item
Let $\varphi\colon Q\to Q$ be a diffeomorphism and
let $\Phi=(T^1_k)^*\varphi$ be the canonical prolongation of $\varphi$
to $(T^1_k)^*Q$. Then:
$$
(i)\; \Phi^*\theta^A=\theta^A \quad , \quad(ii)\,
\Phi^*\omega^A=\omega^A\;.
$$
\item
Let $Z\in\vf (Q)$, and let $Z^{C*}$ be
the canonical prolongation of $Z$ to $(T^1_k)^*Q$. Then
\begin{equation}
\label{inf} (i)\; \Lie(Z^{C*})\theta^A=0 \quad , \quad(ii) \,
\Lie(Z^{C*})\omega^A=0\;.
\end{equation}
\een
\end{lem}
\proof
\ben
\item
Part (i) is a consequence of the commutation rule $\tau_A^*\circ
(T^1_k)^*\varphi=T^*\varphi\circ \tau_A^*$. In fact,
\[\begin{array}{lcl} \left[(T^1_k)^*\varphi\right]^*\theta^A & = &
\left[(T^1_k)^*\varphi\right]^*((\tau_A^*)^*\theta)= [(\tau_A)^*
\circ (T^1_k)^*\varphi]^*\theta=
(T^*\varphi\circ\tau_A^*)^*\theta\\\noalign{\medskip} & =
&(\tau_A^*)^*((T^*\varphi)^*\theta)=(\tau_A^*)^*\theta=\theta^A\;,
\end{array}\]
where we have used that $(T^*\varphi)^*\theta=\theta$ (see
\cite{am}, pag. 180).

Part (ii) is a direct consequence of (i).
\item
 Since the infinitesimal generator of the complete lift $Z^{C*}$ of
$Z$ is the canonical prolongation of the infinitesimal generator
of $Z$, from the first item we conclude that (\ref{inf}) holds.
\een
\qed

\subsubsection{$k$-vector fields}

 Let $M$ be a differentiable manifold. Denote by $T^1_kM$
 the Whitney sum $TM\oplus\stackrel{k}{\dots}\oplus TM$ of $k$
copies of $TM$, with projection
$\tau : T^1_kM \to M$, $\tau({v_1}_q,\ldots , {v_k}_q)=q$.

\begin{definition}
 A  {\rm $k$-vector} field on $M$ is a section ${\bf X} \colon M \longrightarrow T^1_kM$
of the projection $\tau$.
\end{definition}

Since $T^{1}_{k}M$ is  the Whitney sum
$TM\oplus \stackrel{k}{\dots} \oplus TM$ of $k$ copies of $TM$,
we deduce that a $k$-vector field ${\bf X}$ defines
a family of $k$ vector fields $X_{1}, \dots, X_{k}\in\vf(M)$ by
projecting ${\bf X}$ onto every factor; that is,
$X_A=\tau_A\circ{\bf X}$, where
$\tau_A\colon T^1_kQ \rightarrow TQ$ is the
canonical projection on the $A^{th}$-copy $TQ$ of $T^1_kQ$.

\begin{definition}
\label{integsect}
An {\rm integral section}  of the $k$-vector field
${\bf X}=(X_{1}, \dots,X_{k})$, passing through a point $q\in M$, is a map
$\psi\colon U_0\subset \Real^k \rightarrow M$,
defined on some neighborhood  $U_0$ of $0\in \Real^k$,  such that
$$
\psi(0)=q, \, \,
\psi_{*}(t)\left(\frac{\partial}{\partial t^A}\Big\vert_t\right)=X_{A}(\psi (t))
 \quad , \quad \mbox{\rm for every $t\in U_0$, $1\leq A \leq k$}
$$
or, what is equivalent,  $\psi$ satisfies that
$X\circ\psi=\psi^{(1)}$, where  $\psi^{(1)}$ is the first
prolongation of $\psi$  to $T^1_kM$ defined by
$$
\begin{array}{rccl}\label{1prolong}
\psi^{(1)}: & U_0\subset \Real^k & \longrightarrow & T^1_kM
\\\noalign{\medskip}
 & t & \longrightarrow & \psi^{(1)}(t)=j^1_0\psi_t\equiv
 \left(\psi_*(t)\left(\derpar{}{t^1}\Big\vert_t\right),\ldots,
\psi_*(t)\left(\derpar{}{t^k}\Big\vert_t\right)\right) \, .
 \end{array}
$$
 A $k$-vector field ${\bf X}=(X_1,\ldots , X_k)$ on $M$ is
{\rm integrable} if there is an integral section passing through
every point of $M$.
\end{definition}

In local coordinates, we have
\begin{equation}
\label{localfi11}
\psi^{(1)}(t^1, \dots, t^k)=\left( \psi^i (t^1, \dots, t^k),
\frac{\partial\psi^i}{\partial t^A} (t^1, \dots, t^k)\right), \qquad  1\leq A\leq k\, ,\, 1\leq
i\leq n \, .
\end{equation}

\subsection{Hamiltonian formalism: $k$-symplectic Hamiltonian systems}
 \protect\label{kshs}

Let $H \colon (T^1_k)^*Q \to \Real$ be a Hamiltonian function. The
family $((T^1_k)^*Q,\omega^A,H)$ is called a {\sl $k$-symplectic
Hamiltonian system}. The {\sl Hamilton-de Donder-Weyl equations} for
this system are the following set of partial differential equations
\begin{equation}
\label{he}
 \frac{\displaystyle
\partial H}{\displaystyle\partial q^i}\Big\vert_{\psi(t)}=
-\sum_{A=1}^k\frac{\displaystyle \partial\psi^A_i} {\displaystyle
\partial t^A}\Big\vert_t
\quad , \quad \frac{\displaystyle \partial H} {\displaystyle
\partial p^A_i}\Big\vert_{\psi(t)}= \frac{\displaystyle \partial\psi^i}{\displaystyle
\partial t^A}\Big\vert_t, \quad 1\leq i\leq n, \, 1\leq\ A \leq k\, ,
\end{equation}
where $\psi\colon\Real^k\to (T^1_k)^*Q$,
$\psi(t)=(\psi^i(t),\psi^A_i(t))$, is a solution.

We denote by $\vf^k_H((T^1_k)^*Q)$ the set of $k$-vector fields
${\bf X}=(X_1,\dots,X_k)$ on $(T^1_k)^*Q$, which are solutions to the equations
\begin{equation}
\sum_{A=1}^k\inn(X_A)\omega^A=\d H\;.
 \label{generic}
\end{equation}
Then, if ${\bf X}\in\vf^k_H((T^1_k)^*Q)$ is integrable, and $\psi\colon\Real^k\to (T^1_k)^*Q$
is an integral section of ${\bf X}$, then, from (\ref{canosym}),
we obtain that $\psi(t)=(\psi^i(t),\psi^A_i(t))$ is a solution to
the system (\ref{he}).

\subsection{Symmetries and conservation laws}
 \protect\label{ntmvf}

Let $((T^1_k)^*Q,\omega^A,H)$ be a $k$-symplectic Hamiltonian system,
and its associated Hamilton-de Donder-Weyl equations (\ref{he}).

First, following \cite{Olver}, we introduce the next definition :

\begin{definition}
\label{Olver}
A {\rm conservation law} (or a {\rm conserved quantity})
 for the Hamilton-de Donder-Weyl equations (\ref{he})
is a map  ${\cal F}=({\cal F}^1 , \ldots , {\cal F}^k)\colon (T^1_k)^*Q \to \Real^k$
such that the divergence of
$$
{\cal F}\circ \psi=({\cal F}^1 \circ \psi, \ldots , {\cal F}^k \circ
\psi)\colon U_0\subset\Real^k \to \Real^k
$$
 is zero for every solution  $\psi$ to the Hamilton-de Donder-Weyl equations (\ref{he}); that
is
$$
 \sum_{A=1}^k \frac{\partial ({\cal F}^A \circ \psi)}{\partial t^A}=0\;.
$$
\end{definition}

\begin{prop}
If ${\cal F}=({\cal F}^1,\ldots,{\cal F}^k)\colon (T^1_k)^*Q\to \Real^k$
is a conservation law then  for every
integrable $k$-vector field ${\bf X}=(X_1,\dots,X_k)$ in
$\vf^k_H((T^1_k)^*Q)$, we have that
$$
\sum_{A=1}^k\Lie({X_A}){\cal F}^A=0\;.
$$
\end{prop}
\proof
If ${\bf X}=(X_1,\dots,X_k)\in\vf^k_H((T^1_k)^*Q)$
is integrable and $\psi\colon\Real^k\to (T^1_k)^*Q$
is an integral section of ${\bf X}$, then
the following relation holds for every $t\in\Real^k$ and $A=1,\ldots,k$,
$$
X_A(\psi(t))=\psi_*(t)\left(\derpar{}{t^A}\Big\vert_t\right)
$$
and therefore
$$
  \sum_{A=1}^k\Lie(X_A){\cal F}^A=
\sum_{A=1}^k\psi_*(t)\left(\derpar{}{t^A}\Big\vert_t\right)({\cal
F}^A)= \sum_{A=1}^k
 \derpar{({\cal F}^A\circ\psi)}{t^A}\Big\vert_t = 0
$$
since $\psi$ is a solution to the Hamilton-de Donder-Weyl equations (\ref{he}).
\qed

\begin{quote}{\bf Remark}:
The case $k=1$ corresponds to  Classical Mechanics. In
this case we know that ${\cal F}$ is a constant of the motion if
and only if $L(X_H){\cal F}=0$, where $X_H$ is the Hamiltonian
vector field defined by $i(X_H)\omega=dH$.
\end{quote}

 \begin{definition}
 \label{symH}
\ben
\item
A {\rm symmetry} of the $k$-symplectic Hamiltonian system
$((T^1_k)^*Q,\omega^A,H)$ is a diffeomorphism $\Phi\colon (T^1_k)^*Q
\to (T^1_k)^*Q $ such that, for every solution $\psi$ to  the
Hamilton-de Donder-Weyl equations (\ref{he}), we have that
$\Phi\circ\psi$ is also a solution to these equations.

In the particular case that $\Phi=(T^1_k)^*\varphi$ for some
$\varphi\colon Q\to Q$ (i.e.; $\Phi$ is the canonical lifting of
some diffeomorphism in $Q$), the symmetry
$\Phi$ is said to be {\rm natural}.
\item
 An {\rm infinitesimal symmetry}
of the $k$-symplectic Hamiltonian system $((T^1_k)^*Q,\omega^A,H)$
 is a vector field $Y\in\vf((T^1_k)^*Q)$ whose local flows are local symmetries.

In the particular case where $Y=Z^{C*}$ for some
$Z\in\vf (Q)$, (i.e.; $Y$ is the canonical lifting of
some vector field in $Q$), the infinitesimal symmetry
$Y$ is said to be {\rm natural}.
\een
\end{definition}

As a consequence of the definition, all the results that we state for symmetries
also hold for infinitesimal symmetries.

A first straightforward consequence of definitions \ref{Olver} and
\ref{symH}  is:

 \begin{prop}
If $\Phi\colon (T^1_k)^*Q\to (T^1_k)^*Q$ is a symmetry of a
$k$-symplectic Hamiltonian system and  ${\cal F}=({\cal F}^1, \ldots
, {\cal F}^k)\colon (T^1_k)^*Q\to \Real^k$
 is a conservation law, then so is $\Phi^*{\cal F}=(\Phi^*{\cal F}^1,\ldots,\Phi^*{\cal F}^k)$.
\label{generador} \end{prop}

 There is a class of symmetries which play a relevant role
 as generators of conserved quantities:
 \begin{prop}
Let $\Phi\colon (T^1_k)^*Q\to (T^1_k)^*Q$ be a diffeomorphism. If
\[\Phi^*\omega^A=\omega^A\,,\quad 1\leq A \leq k\quad \makebox{ and
}\quad \Phi^*H=H \makebox{  (up to a constant).}
\]
then $\Phi$ is a symmetry of the $k$-symplectic Hamiltonian system
 $((T^1_k)^*Q,\omega^A,H)$.
 \end{prop}
\proof
 We must prove that,
if $\psi\colon U_0\subset \Real^k \to (T^1_k)^*Q$
 is a solution to the Hamilton-de Donder-Weyl equations
(\ref{he}), then $\Phi\circ\psi$ is also a solution, that is,
$$
(a)\quad \frac{\partial H}{\partial
q^i}\Big\vert_{(\Phi\circ\psi)(t)}=
 -\sum_{A=1}^k\ds\frac{\partial (\Phi\circ\psi )^A_i}{\partial t^A}\Big\vert_t
 \quad , \quad (b)\quad
\frac{\partial H}{\partial p^A_i}\Big\vert_{(\Phi\circ\psi)(t)}=
\frac{\partial (\Phi\circ\psi)^i}{\partial t^A}\Big\vert_{t}\;.
$$
In local coordinates, we write the diffeomorphism $\Phi\colon
(T^1_k)^*Q\to (T^1_k)^*Q$ as follows
\[\Phi(q^j,p^B_j)=(\Phi^i(q^j,p^B_j),\Phi^A_i(q^j,p^B_j))\quad .
\]
The condition $\Phi^*\omega^A=\omega^A$ implies \bea \nonumber 0
&=& \derpar{\Phi^i}{q^j}\Big\vert_{w}
\derpar{\Phi^A_i}{q^k}\Big\vert_{w}\;,
\\
\label{omegapp} 0 &=& \derpar{\Phi^i}{p^B_j}\Big\vert_{w}
\derpar{\Phi^A_i}{p^C_k}\Big\vert_{w}\;,
\\
\nonumber \delta^k_j\delta^A_C &=& \derpar{\Phi^i}{q^j}\Big\vert_{w}
\derpar{\Phi^A_i}{p^C_k}\Big\vert_{w} -
\derpar{\Phi^i}{p^C_k}\Big\vert_{w}
\derpar{\Phi^A_i}{q^j}\Big\vert_{w}\;.
 \eea
Furthermore, since $\Phi$ is a diffeomorphism,
$\Phi\circ\Phi^{-1}=Id_{(T^1_k)^*Q}$. Applying the chain rule we
obtain:
 \bea \label{partialqq} \delta^i_k &=&
\derpar{(\Phi\circ\Phi^{-1})^i}{q^k}\Big\vert_{w} =
\derpar{\Phi^i}{q^j}\Big\vert_{\Phi^{-1}(w)}
\derpar{(\Phi^{-1})^j}{q^k}\Big\vert_{w} +
\derpar{\Phi^i}{p^A_j}\Big\vert_{\Phi^{-1}(w)}
\derpar{(\Phi^{-1})^A_j}{q^k}\Big\vert_{w}\,,
\\
 0 &=&
\derpar{(\Phi\circ\Phi^{-1})^i}{p^B_k}\Big\vert_{w} =
\derpar{\Phi^i}{q^j}\Big\vert_{\Phi^{-1}(w)}
\derpar{(\Phi^{-1})^j}{p^B_k}\Big\vert_{w} +
\derpar{\Phi^i}{p^A_j}\Big\vert_{\Phi^{-1}(w)}
\derpar{(\Phi^{-1})^A_j}{p^B_k}\Big\vert_{w}\,,
\nonumber \\
\label{partialpq} 0 &=&
\derpar{(\Phi\circ\Phi^{-1})^A_i}{q^j}\Big\vert_{w} =
\derpar{\Phi^A_i}{q^k}\Big\vert_{\Phi^{-1}(w)}
\derpar{(\Phi^{-1})^k}{q^j}\Big\vert_{w} +
\derpar{\Phi^A_i}{p^B_k}\Big\vert_{\Phi^{-1}(w)}
\derpar{(\Phi^{-1})^B_k}{q^j}\Big\vert_{w}\,,
\\
\label{partialpp} \delta^i_j\;\delta^A_C &=&
\derpar{(\Phi\circ\Phi^{-1})^A_i}{p^C_j}\Big\vert_{w} =
\derpar{\Phi^A_i}{q^k}\Big\vert_{\Phi^{-1}(w)}
\derpar{(\Phi^{-1})^k}{p^C_j}\Big\vert_{w} +
\derpar{\Phi^A_i}{p^B_k}\Big\vert_{\Phi^{-1}(w)}
\derpar{(\Phi^{-1})^B_k}{p^C_j}\Big\vert_{w}\;.
 \eea
 \ From the equations (\ref{omegapp}-\ref{partialpp}) we obtain
\bea
\label{Phiqp} \derpar{\Phi^s}{q^j}\Big\vert_{\Phi^{-1}(w)} =
\delta^A_B\,\derpar{(\Phi^{-1})^A_j}{p^B_s}\Big\vert_{w} &\quad ,
\quad & \delta^A_D\; \derpar{\Phi^s}{p^C_k}\Big\vert_{\Phi^{-1}(w)} =
-\,\delta^A_C \,\derpar{(\Phi^{-1})^k}{p^D_s}\Big\vert_{w}
\\
 \label{Phipp}
\derpar{\Phi^A_s}{q^j}\Big\vert_{\Phi^{-1}(w)} =
-\,\derpar{(\Phi^{-1})^A_j}{q^s}\Big\vert_{w} & \quad , \quad &
\derpar{\Phi^A_s}{p^C_k}\Big\vert_{\Phi^{-1}(w)} =
\delta^A_C\,\derpar{(\Phi^{-1})^k}{q^s}\Big\vert_{w}
\;.\eea
  From the condition $\Phi^*H=H$ written as follows
\[
H(q^j,p^B_j) = (H\circ\Phi)(q^j,p^B_j)
=H(\Phi^i(q^j,p^B_j),\Phi^A_i(q^j,p^B_j))\;,
\]
we obtain,  for every $w\in(T^1_k)^*Q$.
 \bea
 \nonumber  \derpar{H}{q^j}\Big\vert_{w} &=&
\derpar{H}{q^i}\Big\vert_{\Phi(w)} \derpar{\Phi^i}{q^j}\Big\vert_{w}
+ \derpar{H}{p^A_i}\Big\vert_{\Phi(w)}
\derpar{\Phi^A_i}{q^j}\Big\vert_{w}
\\
\label{partialpH} \\\nonumber \derpar{H}{p^A_j}\Big\vert_{w} &=&
\derpar{H}{q^i}\Big\vert_{\Phi(w)}
\derpar{\Phi^i}{p^A_j}\Big\vert_{w} +
\derpar{H}{p^B_i}\Big\vert_{\Phi(w)}
\derpar{\Phi^B_i}{p^A_j}\Big\vert_{w}\ ,
 \eea
Applying the chain rule, by a straightforward computation one proves
$(a)$ as consequence of (\ref{he}), (\ref{omegapp}),
(\ref{partialqq}), (\ref{partialpq}), (\ref{Phipp}) and
(\ref{partialpH}), and  taking into account (\ref{he}),
(\ref{Phiqp}), (\ref{Phipp}) and (\ref{partialpH}), one proves
$(b)$.  \qed

The case $k=1$ corresponds to Classical Mechanics. In this  case the
above result can be found in \cite{mssv}.

   Taking into account this proposition, we introduce the following definitions:
 \begin{definition}
\ben
\item
 A {\rm Cartan ({\it or} Noether) symmetry} of a $k$-symplectic Hamiltonian system
$((T^1_k)^*Q,\omega^A,H)$ is a diffeomorphism
$\Phi\colon (T^1_k)^*Q\to (T^1_k)^*Q$ such that,
\ben
\item
$\Phi^*\omega^A=\omega^A$, for $A=1,\ldots,k$.
\item
$\Phi^*H=H$  (up to a constant).
\een

If $\Phi=(T^1_k)^*\varphi$ for some
$\varphi\colon Q\to Q$, then the Cartan symmetry
$\Phi$ is said to be {\rm natural}.
\item
 An {\rm infinitesimal Cartan ({\it or} Noether) symmetry}
 is a vector field $Y\in\vf((T^1_k)^*Q)$ satisfying that:
\ben
\item
 $\Lie(Y)\omega^A=0$, for $A=1,\ldots,k$.
\item
$\Lie(Y)H=0$.
\een

If $Y=Z^{C*}$ for some $Z\in\vf (Q)$, then the infinitesimal Cartan symmetry
$Y$ is said to be {\rm natural}.
\een
 \label{CNsym}
 \end{definition}

Furthermore, we have that:

 \begin{prop}
 If $\Phi\colon(T^1_k)^*Q\to(T^1_k)^*Q$ is a Cartan symmetry of a $k$-symplectic Hamiltonian system
 $((T^1_k)^*Q,\omega^A,H)$, and ${\bf X}=(X_1,\dots,X_k)\in\vf^k_H((T^1_k)^*Q)$,
then $\Phi_*{\bf X}=(\Phi_*X_1,\dots,\Phi_*X_k)\in\vf^k_H((T^1_k)^*Q)$.
 \end{prop}
 \proof
Let $\Phi\colon (T^1_k)^*Q\to (T^1_k)^*Q$ be a Cartan symmetry.
 For every ${\bf X}=(X_1,\dots,X_k)\in\vf^k_H((T^1_k)^*Q)$ we calculate
$$
\Phi^*[\sum_{A=1}^k\inn(\Phi_*X_A)\omega^A-\d H]=
\sum_{A=1}^k\inn(X_A)(\Phi^*\omega^A)-\d (\Phi^*H)=
\sum_{A=1}^k\inn(X_A)\omega^A-\d H=0
$$
hence, as $\Phi$ is a diffeomorphism, this is equivalent to
demanding that $\ds\sum_{A=1}^k\inn(\Phi_*X_A)\omega^A-\d H=0$, and
therefore $\Phi_*{\bf
X}=(\Phi_*X_1,\dots,\Phi_*X_k)\in\vf^k_H((T^1_k)^*Q)$.
 \qed

In order to state a geometrical version of Noether's theorem for
$k$-symplectic systems, we restrict our study to the infinitesimal
Cartan symmetries.

First, it is immediate to prove that, if $Y_1,Y_2\in\vf((T^1_k)^*Q)$
 are infinitesimal Cartan symmetries, then so is $[Y_1,Y_2]$.

In addition, a highly relevant result is the following:

\begin{prop}
Let $Y\in\vf((T^1_k)^*Q)$ be an infinitesimal Cartan symmetry of a
 $k$-symplectic Hamiltonian system $((T^1_k)^*Q,\omega^A,H)$.
Then, for $A=1,\ldots,k$, and
for every $p\in (T^1_k)^*Q$, there is an open neighbourhood $U_p\ni p$,
such that:
\ben
\item
 There exist $f^A\in\Cinfty(U_p)$, which are unique up to constant functions, such that
\beq \inn(Y)\omega^A=\d f^A, \qquad \mbox{\rm (on $U_p$)}\;.
\label{funo} \eeq
\item
There exist $\zeta^A\in\Cinfty(U_p)$, verifying that
$\Lie(Y)\theta^A=\d\zeta^A$, on $U_p$; and then \beq
f^A=\inn(Y)\theta^A-\zeta^A, \qquad \mbox{\rm (up to a constant
function, on $U_p$)}\,. \label{fdos} \eeq \een \label{structure}
\end{prop}
\proof
\begin{enumerate}
\item
 It is a consequence of the Poincar\'e Lemma and the condition
$$
0=\Lie(Y)\omega^A=\inn(Y)\d\omega^A+\d\inn(Y)\omega^A=\d\inn(Y)\omega^A\;.
$$
 \item
We have that
$$
\d\Lie(Y)\theta^A=\Lie(Y)\d\theta^A=-\Lie(Y)\omega^A=0
$$
and hence $\Lie(Y)\theta^A$ are closed forms. Therefore, by the
Poincar\'e Lemma, there exist $\zeta^A\in\Cinfty(U_p)$, verifying
that $\Lie(Y)\theta^A=\d\zeta^A$, on $U_p$. Furthermore, as
(\ref{funo}) holds on $U_p$, we obtain that
$$
\d\zeta^A=\Lie(Y)\theta^A= \d\inn(Y)\theta^A+\inn(Y)\d\theta^A=
\d\inn(Y)\theta^A-\inn(Y)\omega^A=\d \{\inn(Y)\theta^A-f^A\}
 $$
and thus (\ref{fdos}) holds.
\qed
 \end{enumerate}

\begin{quote}{\bf Remark}:
As a particular case, those Cartan symmetries
$\Phi\colon(T^1_k)^*Q\to(T^1_k)^*Q$ (resp. infinitesimal Cartan
symmetries $Y\in\vf((T^1_k)^*Q)$) verifying that
$\Phi^*\theta^A=\theta^A$ (resp. $\Lie(Y)\theta^A=0$), for
$A=1,\ldots,k$, are usually called {\sl exact}. It is obvious that
natural Cartan symmetries are exact.

Observe that, for exact infinitesimal Cartan symmetries we have that
$f^A=-\inn(Y)\theta^A$.
\end{quote}

 Finally, the classical {\sl Noether's theorem}
 of Hamiltonian mechanics can be generalized to
 $k$-symplectic field theories as follows:

 \begin{teor}
 \label{Nthsec}
{\rm (Noether's theorem):}
 If $Y\in\vf((T^1_k)^*Q)$ is an infinitesimal Cartan symmetry of a
 $k$-symplectic Hamiltonian system $((T^1_k)^*Q,\omega^A,H)$.
 Then, for every $p\in (T^1_k)^*Q$, there is an open
neighborhood $U_p\ni p$ such that the functions
$f^A=\inn(Y)\theta^A-\zeta^A$, $1\leq A\leq k$, define
 a conservation law  $f=(f^1,\ldots,f^k)$.
 \end{teor}
 \proof
Let $Y\in \vf((T^1_k)^*Q)$ with local expression
$ Y=Y^i\ds\frac{\partial}{\partial q^i}+Y^A_i \ds\frac{\partial}{\partial p^A_i}\;, $
  then from (\ref{funo}) we have
$$
Y^i\delta^A_B=\frac{\partial f^A}{\partial p^B_i} \quad , \quad
-Y^A_i=\frac{\partial f^A}{\partial q^i}\quad ; \quad \mbox{\rm (on
$U_p$)}
$$
Let $\psi\colon\Real^k\to (T^1_k)^*Q$ be a solution to (\ref{he}),
then using the last equalities we obtain \beann
\sum_{A=1}^k\frac{\partial (f^A\circ\psi)}{\partial
t^A}\Big\vert_{t}& = &\left(\frac{\partial f^A}{\partial
q^i}\Big\vert_{\psi(t)}\frac{\partial \psi^i}{\partial
t^A}\Big\vert_{t}+ \frac{\partial f^A}{\partial
p^B_i}\Big\vert_{\psi(t)}\frac{\partial \psi^B_i}{\partial
t^A}\Big\vert_{t}\right) =\left(-Y^A_i\frac{\partial
\psi^i}{\partial t^A}\Big\vert_{t}+ Y^i\sum_{A=1}^k\frac{\partial
\psi^A_i}{\partial t^A}\Big\vert_{t}\right)
\\ &=&
 -\left( Y^A_i\frac{\partial H}{\partial p^A_i}+Y^i\frac{\partial H}{\partial q^i}\right)=
-\Lie(Y)H=0\quad.
 \eeann\qed

In the case $k=1$, the above theorem (Noether's Theorem in the
Hamiltonian formalism) can be found in \cite{mssv}.

Furthermore, we have that:

 \begin{teor}
 {\rm (Noether):}
 If  $\,Y\in\vf((T^1_k)^*Q)$ is an infinitesimal Cartan symmetry of a
 $k$-symplectic Hamiltonian system $((T^1_k)^*Q,\omega^A,H)$. Then, for every
${\bf X}=(X_1,\ldots ,X_k)\in\vf^k_H((T^1_k)^*Q)$, we have
 $$
\sum_{A=1}^k\Lie(X_A)f^A=0 \qquad \mbox{\rm (on $U_p$)}\;.
 $$
 \label{Nth}
 \end{teor}
 \proof
 If $Y\in\vf ((T^1_k)^*Q)$ is a Cartan-Noether symmetry, then, on $U_p$,
taking (\ref{funo}) into account we obtain \beann
\sum_{A=1}^k\Lie(X_A)f^A &=& \sum_{A=1}^k(\d\inn(X_A)f^A+\inn(X_A)\d
f^A)= \sum_{A=1}^k\inn(X_A)\inn(Y)\omega^A
\\ &=&
-\inn(Y)\sum_{A=1}^k\inn(X_A)\omega^A=-\inn(Y)\d H=-\Lie(Y)H=0\,.
\eeann
 \qed

Noether's theorem associates conservation laws to Cartan
 symmetries. However, these kinds of symmetries do not exhaust the set of
 symmetries. As is known, in mechanics there are
 symmetries which are not of Cartan type, and
 which also generate conserved quantities
 (see \cite{LMR-99}, \cite{Ra-95}, \cite{Ra-97},
 for some examples). These are the so-called {\sl hidden
 symmetries}. Different attempts have been made
 to extend Noether's theorem in order
 to include these symmetries and the corresponding conserved
 quantities for mechanical systems (see for instance \cite{SC-81})
 and multisymplectic field theories (see \cite{EMR-99}).

\section{ Lagrangian $k$-symplectic case}

\subsection{Geometric elements}

\subsubsection{The tangent bundle of $k^1$-velocities of a manifold.
Canonical structures}

Let $\tau_Q\colon TQ \to Q$ be the tangent bundle of a $Q$. Let us
denote by $T^1_kQ$ the Whitney sum $TQ\oplus\stackrel{k}{\dots}\oplus TQ$
of $k$ copies of $TQ$, with projection $\tau\colon T^1_kQ\to Q$, $\tau ({v_1}_q,\ldots , {v_k}_q)=q$.

$T^1_kQ$ can be identified with the manifold $J^1_0(\Real^k,Q)$ of
the $k^1$-velocities of $Q$; that is,  $1$-jets of maps $\sigma\colon\R^k\to Q$,
with source at $0\in \Real^k$ and with projection map
 $\tau\colon T^1_kQ \to Q$, $\tau (j^1_{0,q}\sigma)=\sigma (0)=q$; that is,
\[
\begin{array}{ccc}
J^1_0(\Real^k,Q) & \equiv & TQ \oplus \stackrel{k}{\dots} \oplus TQ \\
j^1_{0,q}\sigma & \equiv & ({v_1}_q,\ldots , {v_k}_q)
\end{array}
\]
where $q=\sigma (0)$,  and
 ${v_A}_q=\sigma_*(0)\left(\ds\frac{\partial}{\partial t^A}\Big\vert_0\right)$.
The manifold $T^1_kQ$ is called {\sl the tangent bundle of $k^1$-velocities of $Q$} \cite{mor}.

If $(q^i)$ are local coordinates on $U \subseteq Q$ then the
induced local coordinates $(q^i , v^i)$, $1\leq i \leq n$, in
$TU=\tau_Q^{-1}(U)$ are given by
$q^i(v_q)=q^i(q)$, $v^i(v_q)=v_q(q^i)$,
and  the induced local coordinates $(q^i , v_A^i)$, $1\leq i \leq
n,\, 1\leq A \leq k$, in $T^1_kU=\tau^{-1}(U)$ are given by
$$
q^i({v_1}_q,\ldots , {v_k}_q)=q^i(q),\qquad
v_A^i({v_1}_q,\ldots , {v_k}_q)={v_A}_q(q^i) \, .
$$

For a vector $Z_q\in T_qQ$, and for $A=1,\ldots, k$, we define its
{\sl vertical $A$-lift}, $(Z_q)^{V_A}$, at the point
$({v_1}_q,\ldots,{v_k}_q)\in T_k^1Q$, as the vector tangent to the
fiber $\tau^{-1}(q)\subset T_k^1Q$, which is given by
  $$
(Z_q)^{V_A}({v_1}_q,\ldots,v_A)=
\frac{d}{ds}({v_1}_q,\ldots,{v_{A-1}}_q,v_{A_q}+sZ_q,{v_{A+1}}_q,\ldots,{v_k}_q)\vert_{s=0}
\;. $$ In  local coordinates, if $X_q = a^i
\,\ds\frac{\partial}{\partial q^i}\Big\vert_q$, then
\begin{equation}
\label{xa} (Z_q)^{V_A}({v_1}_q,\ldots,{v_k}_q)= a^i
\displaystyle\frac{\partial}{\partial
v^i_A}\Big\vert_{({v_1}_q,\ldots,{v_k}_q)}\ .
\end{equation}

The {\sl canonical $k$-tangent structure} on $T^1_kQ$ is the set
$(S^1,\ldots,S^k)$ of tensor fields  of type $(1,1)$ defined by
 $$
S^A(w_q)(Z_{w_q})= (\tau_*(w_q)(Z_{w_q}))^{V_A}(w_q) \quad , \quad
\mbox{\rm for $w_q\in T^1_kQ$, $Z_{w_q}\in T_{w_q}(T^1_kQ)$; $A=1,
\ldots , k$}\,.$$
 In local  coordinates, from (\ref{xa})  we have
\begin{equation}
\label{localJA} S^A=\frac{\partial}{\partial v^i_A} \otimes \d
q^i\;.
\end{equation}

  The tensors $S^A$ can be regarded as the
$(0,\ldots,0,\stackrel{A}{1},0,\ldots,0)$-lift of the identity
tensor on $Q$ to $T^1_kQ$ defined in \cite{mor}. In the case $k=1$,
$S^1$ is the well-known canonical tangent structure of the tangent
bundle, (see \cite{cra,grif1,grif2,klein}).

Finally, we introduce the {\sl Liouville vector field} $\Delta\in\vf(T^1_kQ)$,
which is the infinitesimal generator of the following flow
 $$
\psi\colon\Real\times T^1_kQ\longrightarrow T^1_kQ  \quad , \quad
  \psi(s,v_{1_{q}}, \ldots ,v_{k_{q}})= (e^s   v_{1_{q}},\ldots , e^s   v_{k_{q}})\, ,
$$
and in local coordinates it has the form
$$
 \Delta =\sum_{A=1}^kv^i_A \derpar{}{v_A^i}\ .
$$

 $\Delta$ is a sum of vector fields $\Delta_1+\ldots+\Delta_k$, where each $\Delta_A$
is the   infinitesimal  generator of the following flow
 \begin{equation}
 \label{psica}
 \psi^A\colon \Real \times T^1_kQ\longrightarrow T^1_kQ  \quad , \quad
\psi^A(s,v_{1_{q}},\ldots ,v_{k_{q}})=(v_{1_{q}},\ldots,v_{{A-1}_q}, e^s v_{A_{q}},v_{A+1_{q}},\ldots,v_{k_{q}})
\end{equation}
and, in  local coordinates, each $\Delta_A$ has the form
\begin{equation}
\label{localca}
 \Delta_A = v^i_A\frac{\partial}{\partial v_A^i}\quad , \quad
\mbox{\rm for $A=1,\ldots,k$ fixed}\, .
\end{equation}

\subsubsection{Complete lift of diffeomorphisms and vector fields from $Q$ to $T^1_kQ$}

Let $\varphi\colon Q\to Q$ be a differentiable map, then
the {\sl canonical prolongation} of $\varphi$ to $ T^1_kQ$ is
the induced map $T^1_k\varphi:T^1_kQ \to  T^1_kQ$  defined by
$T^1_k\varphi(j^1_0\sigma)=j^1_0(\varphi \circ \sigma)$; that is,
for ${v_1}_q,\ldots , {v_k}_q\in T_qQ$, $q\in Q$.
$$
  T^1_k\varphi({v_1}_q,\ldots , {v_k}_q)=
(\varphi_*(q){v_1}_q,\ldots,\varphi_*(q){v_k}_q) \  .
$$

If $Z$ is a vector field on $Q$, with local $1$-parametric group of
transformations $h_s\colon Q \to Q$, then the local $1$-parametric
group of transformations $T^1_k(h_s)\colon T^1_kQ \to T^1_kQ$
generates a vector field $Z^C$ on $T^1_kQ$, which is called the {\sl
complete lift} of $Z$ to $ T^1_kQ$. If where
$Z=Z^i\ds\frac{\partial}{\partial q^i}$, its local expression is
$$
Z^C=Z^i\frac{\partial}{\partial q^i}+v^j_A\frac{\partial Z^k}
{\partial q^j}\frac{\partial}{\partial v^k_A} \,.
$$

Then, we have the following property:

\begin{lem}
\label{lema.2}
 Let $\Phi=T^1_k\varphi:T^1_kQ \to T^1_kQ$ be  the
canonical prolongation of a diffeomorphism $\varphi:Q\to Q$. Then
$$
(a) \quad \Phi_* \circ S^A = S^A \circ \Phi_* \quad , \quad (b)
\quad \Phi_*\Delta_A=\Delta_A \quad , \quad \mbox{\rm  for $A=1,\ldots,k$} \quad .
$$
\end{lem}
\proof
(a) It is a direct consequence of local expression of $S^A$ and
the local expression of $T^1_k\varphi$,
$$
T^1_k\varphi(q^i,v^i_A)=(\varphi^j(q^i),v_A^i\frac{\partial
\varphi^j}{\partial q^i}) \ .
$$

 (b) It is a consequence of $T^1_k\varphi\circ\psi_t^A=\psi_t^A\circ T^1_k\varphi$,
 where $\psi_t^A$ are the local $1$-parameter groups of diffeomorphisms
(\ref{psica}) generated by $\Delta_A$.
\qed

This means that canonical liftings of diffeomorphisms and vector
fields preserve the canonical structures of $T^1_kQ$.

\subsubsection{Second-order partial differential equations in $T^1_kQ$}

The aim of this subsection is to characterize the integrable
$k$-vector fields on $T^1_kQ$ such that their integral sections
are first prolongations $\phi^{(1)}$ of maps  $\phi\colon\Real^k\to Q$.

Remember that a $k$-vector field in $T^1_kQ$ is a section
$\mathbf{\Gamma}\colon T^1_kQ\longrightarrow T^1_k(T^1_kQ)$
of the canonical projection $\tau_{T^1_kQ}\colon T^1_k(T^1_kQ)\to T^1_kQ$. Then:

\begin{definition}
\label{sode0}
A  {\rm second order partial differential equation ({\sc sopde})}
is a $k$-vector field $\mathbf{\Gamma}=(\Gamma_1,\ldots,\Gamma_k)$ in
$T^1_kQ$ which is a section of the projection
$T^1_k\tau\colon T^1_k(T^1_kQ)\rightarrow T^1_kQ$; that is,
$$
T^1_k\tau\circ\mathbf{\Gamma}={\rm Id}_{T^1_kQ} \ ,
$$
 or, what is equivalent,
$$
\tau_*(w_q)(\Gamma_A(w_q))= v_{A_q}\quad ,\quad
\mbox{\rm for $w_q=(v_{1_q},\ldots, v_{k_q})\in T^1_kQ$, $A=1,\ldots , k$}\ .
$$
\end{definition}

In the case $k=1$, this is the definition of a second order
differential equation ({\sc sode}).

 From a direct computation
in local coordinates we obtain that the local expression of a {\sc sopde}
$\mathbf{\Gamma}=(\Gamma_1,\ldots,\Gamma_k) $ is
\begin{equation}
\label{localsode1}
\Gamma_A(q^i,v^i_A)= v^i_A\frac{\partial} {\partial q^i}+
(\Gamma_A)^i_B \frac{\partial} {\partial v^i_B}\quad ,\quad
 1\leq A \leq k \quad , \quad
(\Gamma_A)^i_B\in\Cinfty(T^1_kQ)\;.
\end{equation}

If $\psi\colon\Real^k \to T^1_kQ$ is an integral section of
$\mathbf{\Gamma}=(\Gamma_1,\ldots,\Gamma_k)$, locally given by
$\psi(t)=(\psi^i(t),\psi^i_B(t))$, then from Definition
\ref{integsect} and (\ref{localsode1}) we deduce
\begin{equation}
\label{solsopde}
 \frac{\partial\psi^i} {\partial t^A}\Big\vert_t=\psi^i_A(t)\quad ,\quad
\frac{\partial\psi^i_B} {\partial t^A}\Big\vert_t=(\Gamma_A)^i_B(\psi(t))\, .
\end{equation}

 From (\ref{localfi11}) and (\ref{solsopde}) we obtain the following proposition.

\begin{prop}
 \label{sope1}
Let $\mathbf{\Gamma}=(\Gamma_1,\ldots,\Gamma_k)$ be an integrable
{\sc sopde}. If $\psi$ is an integral section of ${\bf \Gamma}$
then $\psi=\phi^{(1)}$, where $\phi^{(1)}$ is the first
prolongation of the map
$\phi=\tau\circ\psi\colon\Real^k\stackrel{\psi}{\to}T^1_kQ\stackrel{\tau}{\to}Q$,
and $\phi$ is a solution to the system of second order partial
differential equations
\begin{equation}
\label{nn1}
 \frac{\partial^2 \phi^i}{\partial t^A\partial t^B}(t)=
(\Gamma_A)^i_B\left(\phi^i(t),\frac{\partial\phi^i}{\partial
t^C}(t)\right) \quad  1\leq i\leq n\, ; 1\leq A,B\leq k.
\end{equation}
Conversely, if $\phi\colon\Real^k \to Q$ is any map satisfying
(\ref{nn1}), then $\phi^{(1)}$ is an integral section of
$\mathbf{\Gamma}=(\Gamma_1,\ldots,\Gamma_k)$.
\end{prop}

 From (\ref{nn1}) we deduce that if $\mathbf{\Gamma}$ is an
integrable {\sc sopde} then $(\Gamma_A)^i_B=(\Gamma_B)^i_A$ for
all $A,B=1,\ldots, k$.

The following characterization of {\sc sopde}s can be given using
the canonical $k$-tangent structure of $T^1_kQ$ (see
(\ref{localJA}), (\ref{localca}) and (\ref{localsode1})):

\begin{prop}
    \label{pr235}
A $k$-vector field $\mathbf{\Gamma}=(\Gamma_1,\ldots,\Gamma_k)$ on
$T^1_kQ$ is a {\sc sopde}  if, and only if,
$S^A(\Gamma_A)=\Delta_A$, for all $A: 1 \ldots , k$\ .
\end{prop}

\subsection{Lagrangian formalism: $k$-symplectic Lagrangian systems}
 \protect\label{ksls}

  In Classical Mechanics, the  symplectic structure of Hamiltonian
theory and the tangent structure of Lagrangian theory play
complementary roles (see Refs. [13,15,16]).
In this subsection, we recall  the Lagrangian formalism developed
by G\"unther \cite{gun} using  the polysymplectic structures. Here
we can see how the polysymplectic structures and the $k$-tangent structures
 also play a complementary role in field theory.

  Let $L\colon T^1_kQ  \to \R$ be a Lagrangian.
The {\sl generalized Euler-Lagrange equations} for $L$ are:
\begin{equation}
\label{ELe} \sum_{A=1}^k\frac{\partial}{\partial t^A}\Big\vert_t
\left(\frac{\displaystyle\partial L}{\partial
v^i_A}\Big\vert_{\psi(t)}\right)= \frac{\partial L}{\partial
q^i}\Big\vert_{\psi(t)}
 \quad , \quad
v^i_A(\psi(t))= \frac{\partial\psi^i}{\partial t^A}
\end{equation}
whose solutions are maps $\psi\colon\Real^k \to T^1_kQ$. Let us observe
that $\psi(t)=\phi^{(1)}(t)$, for some $\phi=\tau \circ  \psi$.

 We introduce a family of $1$-forms  $\theta_L^A$ on $T^1_kQ$,
$1\leq A \leq k$,
  using the $k$-tangent structure, as follows
\begin{equation}\label{betaloc}
 \theta_L^A=  dL \circ S^A  \, \quad 1 \leq
A \leq k  \quad ,
\end{equation}
and hence we define $\omega_L^A=-\d\theta_L^A$.

In local natural coordinates we have \bea \label{thetala} \theta_L^A
&=& \ds\frac{\partial L}{\partial v^i_A}\, dq^i
 \\
\label{omegala}
\omega_L^A &=&
\d q^i \wedge \d\left(\frac{\partial L}{\partial v^i_A}\right)=
\frac{\partial ^2 L}{\partial q^j\partial v^i_A}\d q^i\wedge\d q^j +
\frac{\partial ^2 L}{\partial v^j_B\partial v^i_A}\d q^i\wedge\d v^j_B \ .
\eea

We also introduce the {\sl Energy lagrangian function}
$E_L=\Delta(L)-L\in\Cinfty(T^1_kQ)$, whose local expression is
\beq
E_L=v^i_A\frac{\partial L}{\partial v^i_A}-L \ .
\label{energyL}
\eeq

Then, the family $(T^1_kQ,\omega_L^A,E_L)$ is
called a {\sl $k$-symplectic Lagrangian system}.

\begin{definition}
 The Lagrangian $L\colon T^1_kQ\longrightarrow \Real$ is
said to be regular if the matrix
 $\left(\frac{\partial^2 L}{\partial v^i_A \partial v^j_B}\right)$ is not singular
at every point of $T^1_kQ$.
\end{definition}

\begin{quote}{\bf Remark}:
Let us observe that the condition $L$ regular is equivalent to
$(\omega_L^1,\ldots, \omega_L^k)$ being a polysympletic form and
$(\omega_L^1,\ldots, \omega_L^k;V)$, where $V=Ker \tau_*$, is a
$k$-symplectic structure (see \cite{fam}).
\end{quote}

This $k$-symplectic (polysymplectic) structure, associated to $L$,
was also introduced by G\"{u}nther \cite{gun} using the Legendre
transformation.

The \emph{Legendre map} $FL\colon T^1_kQ\to (T^1_k)^*Q$ was
introduced by G\"{u}nther, \cite{gun} and was rewritten in \cite{fam}
as follows: if $(v_{1_q},\ldots, v_{k_q})\in (T^1_k)_q Q$,
\[[FL(v_{1_q},\ldots, v_{k_q})]^A(u_q) = \frac{d}{ds}\Big\vert_{s=0}L
(v_{1_q},\ldots, v_{A_q}+su_q,\ldots, v_{k_q})\;, \] for each
$A=1,\ldots, k$ and $u_q\in T_qQ$. Locally $FL$ is given by
\begin{equation}\label{locFL}
FL(q^i,v^i_A)=(q^i,\ds\derpar{L}{v^i_A})\;.\end{equation}

In fact, form (\ref{thetala}) and (\ref{locFL}),we easily obtain the
following Lemma.
\begin{lem} For every $1\leq A\leq k\,,\quad
\omega_L^A=(FL)^*\omega^A$, where $(\omega^1,\ldots, \omega^k)$ are
the $2$-forms of the canonical polysymplectic structure.
\end{lem}
Then, from (\ref{locFL}) we obtain the following Proposition.
\begin{prop}Let $L$ be a Lagrangian. The following conditions are
equivalent: $(1)$  $\;L$ is regular. $(2)$ $\; FL$ is a local
diffeomorphism. $(3)$ $\;(\omega_L^1,\ldots, \omega_L^k)$ is a
polysimplectic structure on $T^1_kQ$.
\end{prop}

 As in the Hamiltonian case,
consider a $k$-symplectic Lagrangian system $(T^1_kQ,\omega_L^A,E_L)$, and
denote by $\vf^k_L(T^1_kQ)$ the set of $k$-vector fields
${\bf \Gamma}=(\Gamma_1,\dots,\Gamma_k)$ in $T^1_kQ$, which are solutions
to the equation
 \begin{equation}
\label{genericEL} \sum_{A=1}^k \inn(\Gamma_A)\omega_L^A=\d E_L\, .
 \end{equation}

If each $ \Gamma _A$ is locally given by
$$
\Gamma _A  =  ( \Gamma _A)^i \frac{\partial}{\partial  q^i} + (
\Gamma _A)^i_B\frac{\partial}{\partial v^i_B}\quad ,
$$
 then ${\bf \Gamma}=( \Gamma _1, \ldots , \Gamma_k)$ is a solution to (\ref{genericEL})
 if, and only if, $( \Gamma_A)^i$ and $( \Gamma _A)^i_B$ satisfy the system of equations
\beann
  \left( \frac{\partial^2 L}{\partial q^i \partial v^j_A} - \frac{\partial^2 L}{\partial q^j \partial v^i_A}
\right) \, ( \Gamma _A)^j - \frac{\partial^2 L}{\partial v_A^i
\partial v^j_B} \, ( \Gamma _A)^j_B &=&
 v_A^j \frac{\partial^2 L}{\partial q^i\partial v^j_A} - \frac{\partial  L}{\partial q^i } \, ,
\\
\frac{\partial^2 L}{\partial v^j_B\partial v^i_A} \, ( \Gamma_A)^i
 &=& \frac{\partial^2 L}{\partial v^j_B\partial v^i_A} \, v_A^i \quad .
\eeann

If the Lagrangian is regular, the above equations  are equivalent
to the equations
\bea
\label{locel4}
\frac{\partial^2 L}{\partial q^j \partial v^i_A} v^j_A +
\frac{\partial^2 L}{\partial v_A^i\partial v^j_B}( \Gamma _A)^j_B =
\frac{\partial  L}{\partial q^i}
\\
\label{locel3}
 ( \Gamma _A)^i= v_A^i\quad , \quad 1\leq i \leq n, \,  1\leq A \leq k\ .
\eea

Thus, if $L$ is a regular Lagrangian,  we deduce:
\begin{itemize}
\item
 If ${\bf  \Gamma }=( \Gamma _1,\dots, \Gamma _k)$ is a
solution to (\ref{genericEL}) then it is a {\sc sopde}, (see (\ref{locel3})).
\item
 Equation (\ref{locel4}) leads to define
local solutions to (\ref{genericEL}) in a neighborhood of each
point of $T^1_kQ$ and, using a partition of unity, global
solutions to (\ref{genericEL}).
 \item
  Since ${\bf \Gamma }=(\Gamma _1,\dots, \Gamma _k)\in\vf^k_L(T^1_kQ)$ is a {\sc sopde},
 from Proposition \ref{sope1} we know that, if it is integrable, then its integral
sections are first prolongations $\phi^{(1)}\colon\Real^k\to T^1_kQ$ of
maps $\phi\colon \Real^k \to Q$, and from (\ref{locel4}) we deduce that
$\phi$ is a solution to the Euler-Lagrange equations (\ref{ELe}).
 \item
In the case $k=1$,
 the equation (\ref{genericEL}) is $\imath_{\Gamma} \omega_L = dE_L$,
which is the dynamical equation of the Lagrangian formalism in Mechanics.
\end{itemize}

Throughout this paper, we only consider regular Lagrangians.

\subsection{Symmetries and conservation laws}
 \protect\label{clgcs}

Of course, regarding these topics,
of course, all the definitions stated in Section \ref{ntmvf}
for the Hamiltonian case are applied to the Lagrangian case,
just considering $(T^1_kQ,\omega_L^A,E_L)$
as a Hamiltonian system with Hamiltonian function $E_L$.
In particular, we can define:

 \begin{definition}
\label{ley cons} A map ${\cal F}=({\cal F}^1 , \ldots , {\cal
F}^k)\colon T^1_kQ\to \Real^k$ is a {\rm conservation law} (or a
{\rm conserved quantity}) for the Euler-Lagrange equations
(\ref{ELe}) if the divergence of ${\cal F}\circ\phi=({\cal
F}^1\circ\phi^{(1)},\ldots,{\cal
F}^k\circ\phi^{(1)})\colon\Real^k\to\Real^k$ is zero, for every
$\phi\colon\Real^k\to Q$ solution to the Euler-Lagrange equations
(\ref{ELe}); that is
$$
\sum_{A=1}^k\frac{\partial ({\cal F}^A \circ \phi^{(1)})}{\partial
t^A}=0\;.
$$
 \end{definition}

Therefore, if ${\cal F}=({\cal F}^1,\ldots,{\cal F}^k)\colon
T^1_kQ\to \Real^k$ is a conservation law then,  for every integrable
$k$-vector field ${\bf\Gamma}=(\Gamma_1,\dots,\Gamma_k)$ in
$\vf^k_L(T^1_kQ)$, we have that
$$
\sum_{A=1}^k\Lie({\Gamma_A}){\cal F}^A=0\;.
$$

\begin{definition}
 \label{symL}
\ben
\item
A {\rm symmetry} of the $k$-symplectic Lagrangian system
$(T^1_kQ,\omega_L^A,E_L)$ is a diffeo\-morphism $\Phi\colon T^1_kQ
\to T^1_kQ $ such that, for every solution $\phi$ to  the
Euler-Lagrange equations (\ref{ELe}), we have that
$\Phi\circ\phi^{(1)}=\rho^{(1)}$,
 where $\rho\colon\R^k\to Q$ is also a solution to these equations.

In the particular case that $\Phi=T^1_k\varphi$ for some
$\varphi\colon Q\to Q$ (i.e.; $\Phi$ is the canonical lifting of
some diffeomorphism in $Q$), the symmetry
$\Phi$ is said to be {\rm natural}.
\item
 An {\rm infinitesimal symmetry}
of the $k$-symplectic Lagrangian system $(T^1_kQ,\omega_L^A,E_L)$
 is a vector field $Y\in\vf(T^1_kQ)$ whose local flows are local symmetries.

In the particular case that $Y=Z^C$ for some
$Z\in\vf (Q)$, (i.e.; $Y$ is the canonical lifting of
some vector field in $Q$), the infinitesimal symmetry
$Y$ is said to be {\rm natural}.
\een
\end{definition}

As in the Hamiltonian case, we have that:

\begin{prop}
Let $\Phi\colon T^1_kQ\to T^1_kQ$ be a diffeomorphism. If $\Phi$
satisfies
\[\Phi^*\omega_L^A=\omega_L^A\,,\quad 1\leq A \leq k\quad \makebox{ and
}\quad \Phi^*E_L=E_L \makebox{  (up to a constant).}
\]
then $\Phi$ is a symmetry of the $k$-symplectic Lagrangian system
 $(T^1_kQ,\omega_L^A,E_L)$.
 \end{prop}
 \proof We must prove that, if $\phi\colon U_0\subset\Real^k \to Q$
is a solution to the Euler-Lagrange equations (\ref{ELe}), then
$\Phi\circ\phi^{(1)}$ is also a solution. However, it is
well-known that this is equivalent to proving that
$FL\circ\Phi\circ\phi^{(1)}\colon U_0\subset\Real^k\to (T^1_k)^*Q$
is a solution to the Hamilton-de Donder-Weyl equations, (\ref{he});
that is \begin{eqnarray*}
(a)\quad\derpar{H}{p^A_i}\Big\vert_{(FL\circ\Phi\circ\phi^{(1)})(t)}
&=& \derpar{(FL\circ\Phi\circ\phi^{(1)})^i}{t^A}\Big\vert_{t}
\\(b)\quad
\derpar{H}{q^i}
\Big\vert_{(FL\circ\Phi\circ\phi^{(1)})(t)}\hspace{0,1cm} &=&
-\ds\sum_{A=1}^k \derpar{(FL\circ\Phi\circ\phi^{(1)})^i_A}{t^A}\Big
\vert_{t}\;,
\end{eqnarray*} with Hamiltonian $H=E_L\circ FL^{-1}$.

Let us suppose that $\Phi:T^1_kQ\to T^1_kQ$, locally given by
$\Phi(q^j,v^j_B)=(\Phi^i(q^j,v^j_B),\Phi^i_A(q^j,v^j_B))$ satisfies
the conditions $\Phi^*\omega_L^A= \omega_L^A$ and $E_L=\Phi^*E_L$.

In order to prove $(a)$ and $(b)$ we will use four groups of
identities.
 From the condition $\Phi^*\omega_L^A= \omega_L^A$ we obtain the  first group of identities:
  for everY $w\in T^1_kQ$,
\begin{eqnarray}
\nonumber\derpars{L}{q^j}{v^i_A}\Big\vert_{w} &=& \left(
\derpars{L}{q^k}{v^l_A}\Big\vert_{\Phi(w)}
\derpar{\Phi^k}{q^j}\Big\vert_{w} +
\derpars{L}{v^k_C}{v^l_A}\Big\vert_{\Phi(w)}
\derpar{\Phi^k_C}{q^j}\Big\vert_{w}
\right)\derpar{\Phi^l}{q^i}\Big\vert_{w}\;,
\\ \nonumber
\derpars{L}{v^j_B}{v^i_A}\Big\vert_{w} &=& \left(
\derpars{L}{q^k}{v^l_A}\Big\vert_{\Phi(w)}
\derpar{\Phi^k}{v^j_B}\Big\vert_{w} +
\derpars{L}{v^k_C}{v^l_A}\Big\vert_{\Phi(w)}
\derpar{\Phi^k_C}{v^j_B}\Big\vert_{w}
\right)\derpar{\Phi^l}{q^i}\Big\vert_{w}
\\ &-& \label{omegaLqv}  \left(
\derpars{L}{q^k}{v^l_A}\Big\vert_{\Phi(w)}
\derpar{\Phi^k}{q^i}\Big\vert_{w} +
\derpars{L}{v^k_C}{v^l_A}\Big\vert_{\Phi(w)}
\derpar{\Phi^k_C}{q^i}\Big\vert_{w} \right)
\derpar{\Phi^l}{v^j_B}\Big\vert_{w}\;,
\\\nonumber 0 &=& \left(
\derpars{L}{q^k}{v^l_A}\Big\vert_{\Phi(w)}
\derpar{\Phi^k}{v^j_B}\Big\vert_{w} +
\derpars{L}{v^k_C}{v^l_A}\Big\vert_{\Phi(w)}
\derpar{\Phi^k_C}{v^j_B}\Big\vert_{w} \right)
\derpar{\Phi^l}{v^m_D}\Big\vert_{w}\;.
\end{eqnarray}
Applying the chain rule to $\Phi\circ\Phi^{-1}=Id_{T^1_kQ}$, we have
the second group.
\begin{eqnarray} \label{partqq} \delta^i_k &=&
\derpar{\Phi^i}{q^j}\Big\vert_{\Phi^{-1}(w)}
\derpar{(\Phi^{-1})^j}{q^k}\Big\vert_{w} +
\derpar{\Phi^i}{v^j_A}\Big\vert_{\Phi^{-1}(w)}
\derpar{(\Phi^{-1})^j_A}{q^k}\Big\vert_{w}\;,
\\
\label{partqv} 0 &=& \derpar{\Phi^i}{q^j}\Big\vert_{\Phi^{-1}(w)}
\derpar{(\Phi^{-1})^j}{v^k_B}\Big\vert_{w} +
\derpar{\Phi^i}{v^j_A}\Big\vert_{\Phi^{-1}(w)}
\derpar{(\Phi^{-1})^j_A}{v^k_B}\Big\vert_{w}\;,
\\
\label{partvq} 0 &=& \derpar{\Phi^i_A}{q^k}\Big\vert_{\Phi^{-1}(w)}
\derpar{(\Phi^{-1})^k}{q^j}\Big\vert_{w} +
\derpar{\Phi^i_A}{v^k_B}\Big\vert_{\Phi^{-1}(w)}
\derpar{(\Phi^{-1})^k_B}{q^j}\Big\vert_{w}\;.
 \end{eqnarray}

 The third group of identities is a consequence of the following fact: if
  $\phi\colon U_0\subset\Real^k\to Q$ is a solution to
Euler-Lagrange's equations, we know that $FL\circ\phi^{(1)}\colon
U_0\subset\Real^k\to (T^1_k)^*Q$ is a solution to  Hamilton-de
Donder-Weyl's equations (\ref{he}). Then from the local expression
of $FL$, (\ref{locFL}) we deduce the following equations.
\begin{eqnarray}
\nonumber\derpar{H}{p^A_i}\Big\vert_{(FL\circ\phi^{(1)})(t)} &=&
\derpar{(FL\circ\phi^{(1)})^i}{t^A}\Big\vert_{t} =
\derpar{\phi^i}{t^A}\Big\vert_{t}\;,
\\\noalign{\medskip}
\derpar{H}{q^i}\Big\vert_{(FL\circ\phi^{(1)})(t)} &=& -
\ds\sum_{A=1}^k\derpar{(FL\circ\phi^{(1)})^i_A}{t^A}\Big\vert_{t}
\nonumber \\ & =& -
\derpars{L}{q^j}{v^i_A}\Big\vert_{\phi^{(1)}(t)}
\derpar{\phi^j}{t^A}\Big\vert_{t}
-\derpars{L}{v^j_B}{v^i_A}\Big\vert_{\phi^{(1)}(t)}
\derpars{\phi^j}{t^A}{t^B}\Big\vert_{t}\;.
\label{HqFL}
\end{eqnarray}

Since $E_L=\Phi^*E_L$ is equivalent to $FL^*H=(FL\circ\Phi)^*H$, by
applying the chain rule again and by using the local expression of
$FL$ (\ref{locFL}), we obtain the last family of identities
\begin{eqnarray} 
&& \derpar{H}{q^i}\Big\vert_{FL(w)} +
\derpar{H}{p^B_j}\Big\vert_{FL(w)}\derpars{L}{q^i}{v^j_B}\Big\vert_{w} =
 \derpar{H}{q^j}\Big\vert_{(FL\circ\Phi)(w)}\derpar{\Phi^j}{q^i}\Big\vert_{w}
\nonumber \\ &&
 +\derpar{H}{p^B_j}\Big\vert_{(FL\circ\Phi)(w)} \left(
\derpars{L}{q^k}{v^j_B}\big\vert_{\Phi(w)}
\derpar{\Phi^k}{q^i}\Big\vert_{w} +
\derpars{L}{v^k_A}{v^j_B}\Big\vert_{\Phi(w)}
\derpar{\Phi^k_A}{q^i}\Big\vert_{w} \right)
\label{Hq}
\\\noalign{\medskip} && \derpar{H}{p^B_j}\Big\vert_{FL(w)}
\derpars{L}{v^i_A}{v^j_B}\Big\vert_{w}=
\derpar{H}{q^j}\Big\vert_{(FL\circ\Phi)(w)}
\derpar{\Phi^j}{v^i_A}\Big\vert_{w}
\nonumber \\ & &  +
\derpar{H}{p^B_j}\Big\vert_{(FL\circ\Phi)(w)} \left(
\derpars{L}{q^k}{v^j_B}\big\vert_{\Phi(w)}
\derpar{\Phi^k}{v^i_A}\Big\vert_{w} +
\derpars{L}{v^k_C}{v^j_B}\Big\vert_{\Phi(w)}
\derpar{\Phi^k_C}{v^i_A}\Big\vert_{w} \right)\;. 
\label{Hv}
\end{eqnarray}
These identities (\ref{Hq}) and (\ref{Hv}) are fundamental
to proof of this proposition. Let us observe that in these
identities we find the partial derivatives
$\ds\derpar{H}{q^i}\Big\vert_{(FL\circ\Phi\circ\phi^{(1)})(t)}$ and
$\ds\derpar{H}{p^A_i}\Big\vert_{(FL\circ\Phi\circ\phi^{(1)})(t)}$,
 which we are searching for, and their relation with the other
partial derivatives
$\ds \derpar{H}{q^i}\Big\vert_{(FL\circ\phi^{(1)})(t)}$ and
$\ds\derpar{H}{p^A_i}\Big\vert_{(FL\circ\phi^{(1)})(t)}$,
wich we know from (\ref{HqFL}).

By a straightforward computation, from equations (\ref{omegaLqv}-
\ref{partqv}), (\ref{HqFL}-\ref{Hv}) one proves that
$$
0=\derpars{L}{v^s_D}{v^l_A}\Big\vert_{(\Phi\circ\phi^{(1)})(t)} \left(
\derpar{H}{p^A_l}\Big\vert_{(FL\circ\Phi\circ\phi^{(1)})(t)}
-\derpar{\Phi^l}{q^j}\Big\vert_{\phi^{(1)}(t)}
\derpar{\phi^j}{t^A}\Big\vert_{t} -
\derpar{\Phi^l}{v^j_B}\Big\vert_{\phi^{(1)}(t)}
\derpars{\phi^j}{t^A}{t^B}\Big\vert_{t}\right)
$$
 and since
$L$ is regular, from the above identity  we deduce that
\begin{equation}\label{regular}
\derpar{H}{p^A_l}\Big\vert_{(FL\circ\Phi\circ\phi^{(1)})(t)}
 = \derpar{\Phi^l}{q^j}\Big\vert_{\phi^{(1)}(t)}
\derpar{\phi^j}{t^A}\Big\vert_{t} +
\derpar{\Phi^l}{v^j_B}\Big\vert_{\phi^{(1)}(t)}
\derpars{\phi^j}{t^A}{t^B}\Big\vert_{t}\;. \end{equation}
Furthermore we have
 \begin{equation}\label{tesis1b}
\derpar{(FL\circ\Phi\circ\phi^{(1)})^l}{t^A}\Big\vert_{t}=
\derpar{\Phi^l}{q^j}\Big\vert_{\phi^{(1)}(t)}
\derpar{\phi^j}{t^A}\Big\vert_{t} +
\derpar{\Phi^l}{v^j_B}\Big\vert_{\phi^{(1)}(t)}
\derpars{\phi^j}{t^A}{t^B}\Big\vert_{t}\end{equation} and thus from
(\ref{regular}) and (\ref{tesis1b}) we obtain the first group,
$(a)$, of the Hamilton-de Donder-Weyl equations.

Finally, from (\ref{omegaLqv}), (\ref{partqq}),
(\ref{partvq}-\ref{Hv}) and (\ref{tesis1b}), by a straightforward
computation, one obtains
\begin{eqnarray}
  & &\ds\sum_{A=1}^k
\derpar{(FL\circ\Phi\circ\phi^{(1)})^m_A}{t^A}\Big \vert_{t} =
-\derpar{H}{q^m}\Big\vert_{(FL\circ\Phi\circ\phi^{(1)})(t)}
\nonumber \\\noalign{\medskip} & & 
+\derpars{L}{q^m}{v^i_A}\Big\vert_{(\Phi\circ\phi^{(1)})(t)}
\left(\derpar{(FL\circ\Phi\circ\phi^{(1)})^i}{t^A}\Big\vert_{t}
-\derpar{H}{p^A_i}\Big\vert_{(FL\circ\Phi\circ\phi^{(1)})(t)}\right)
\label{for b}
\end{eqnarray}
and since  we have already proved $(a)$, from
 (\ref{for b}) and $(a)$ one obtains $(b)$. \qed

Taking into account this proposition, we introduce the following
definitions.
\begin{definition}
\label{nls}
 \ben \item
 A {\rm Cartan (or Noether) symmetry}
 of the $k$-symplectic Lagrangian system $(T^1_kQ,\omega_L^A,E_L)$ is a diffeomorphism
$\Phi\colon T^1_kQ\to T^1_kQ$ such that,
 \ben
 \item
$\Phi^*\omega_L^A=\omega_L^A$, for $A=1,\ldots,k$.
 \item
$\Phi^*E_L=E_L$  (up to a constant).
 \een

If $\Phi=T^1_k\varphi$ for some
$\varphi\colon Q\to Q$, then the Cartan symmetry
$\Phi$ is said to be {\rm natural}.
\item
 An {\rm infinitesimal Cartan (or Noether) symmetry}
 of the $k$-symplectic Lagrangian system $(T^1_kQ,\omega_L^A,E_L)$
 is a vector field $Y\in\vf(T^1_kQ)$ satisfying that:
\ben
\item
 $\Lie(Y)\omega_L^A=0$, for $A=1,\ldots,k$.
\item
$\Lie(Y)E_L=0$.
\een

If $Y=Z^C$ for some $Z\in\vf (Q)$, then the infinitesimal Cartan symmetry
$Y$ is said to be {\rm natural}.
\een
 \end{definition}

\begin{prop}
\label{condnoeth}
Let $Y\in\vf(T^1_kQ)$ be an infinitesimal  Cartan symmetry
 of a  $k$-symplectic Lagrangian system $(T^1_kQ,\omega_L^A,E_L)$. Then,
for $A=1,\ldots,k$, and for every $p\in (T^1_k)Q$, there is an
open neighborhood $U_p\ni p$, such that:
 \ben
 \item
 There exist $f^A\in\Cinfty(U_p)$, which are unique up to constant functions, such that
\begin{equation} \inn(Y)\omega_L^A=\d f^A, \qquad \mbox{\rm (on
$U_p$)}\;. \label{0funo} \end{equation} \item There exist
$\zeta^A\in\Cinfty(U_p)$, verifying that
$\Lie(Y)\theta_L^A=\d\zeta^A$, on $U_p$; and then
\begin{equation}
 f^A=\inn(Y)\theta_L^A-\zeta^A, \qquad \mbox{\rm (up to a constant function, on
 $U_p$)}\;.
 \label{0fdos}
 \end{equation}
\een
\end{prop}
\proof
This is the same proof as in Proposition \ref{structure}.
\qed

Now we can state the version of Noether's Theorem for
infinitesimal Cartan Lagrangian symmetries.

\begin{teor}
 \label{0Nthsec}
 {\rm (Noether's theorem):}
 Let $Y\in\vf(T^1_kQ)$ be an infinitesimal Cartan symmetry
 of a  $k$-symplectic Lagrangian system $(T^1_kQ,\omega_L^A,E_L)$,
  then  for every $p\in T^1_kQ$, there is an
open neighborhood $U_p\ni p$, such that the functions
$f^A=\inn(Y)\theta_L^A-\zeta^A\,,\; 1\leq A\leq k$,
 define a conservation law $f=(f^1,\ldots,f^k)$.
 \end{teor}
\proof
Let $Y\in \vf((T^1_k)^*Q)$ be an infinitesimal  Cartan
 symmetry, with local expression
$$
Y=Y^i \frac{\partial}{\partial q^i}+Y^i_A\
\frac{\partial}{\partial v_A^i}\;.
$$
  Then from (\ref{0funo}), as $Y$ is an infinitesimal Cartan symmetry we have that
\begin{equation}\label{cond1}
\left(\frac{\partial^2 L}{\partial q^k\partial
v^i_A}-\frac{\partial^2 L}{\partial q^i\partial v^k_A}\right)Y^i -
Y^i_B\frac{\partial^2 L}{\partial v^i_B\partial
v^k_A}=\frac{\partial f^A}{\partial q^k}\;
\end{equation}
\begin{equation}\label{cond2}
\frac{\partial^2 L}{\partial v^k_B\partial
v^i_A}Y^i=\frac{\partial f^A}{\partial v^k_B}\;.
\end{equation}
Therefore, since $Y$ is an infinitesimal symmetry, from
$\Lie(Y)E_L=0$ we obtain
\begin{equation}\label{cond3}
Y^i\frac{\partial L}{\partial q^i} =v^k_B\left(Y^i\frac{\partial
^2 L}{\partial q^i\partial v^k_B}+  Y^i_A\frac{\partial^2
L}{\partial v^i_A\partial v^k_B}\right)\;.
\end{equation}
Let $\phi:\Real^k\to Q$ be a solution to the Euler-Lagrange
equations, then from (\ref{ELe}), (\ref{cond1}), (\ref{cond2}) and
(\ref{cond3}) we obtain \beann \sum_{A=1}^k\frac{\partial
(f^A\circ\phi^{(1)})}{\partial t^A}\Big\vert_{t}&=&\sum_{A=1}^k
\left(\ds\frac{\partial f^A}{\partial q^k}\Big\vert_{\phi^{(1)}(t)}
\frac{\partial\phi^k}{\partial t^A}\Big\vert_{t} + \frac{\partial
f^A}{\partial v^k_B}\Big\vert_{\phi^{(1)}(t)}
\frac{\partial^2\phi^k}{\partial t^A\partial
t^B}\Big\vert_{t}\right)
\\&=&\sum_{A=1}^k
Y^i(\phi^{(1)}(t))\left(\frac{\partial^2L}{\partial q^k\partial
v^i_A}\Big\vert_{\phi^{(1)}(t)} \frac{\partial \phi^k}{\partial
t^A}\Big\vert_{t}+ \frac{\partial^2 L}{\partial v^k_B\partial
v^i_A}\Big\vert_{\phi^{(1)}(t)} \frac{\partial^2\phi^k}{\partial
t^A\partial t^B}\Big\vert_{t}\right)
\\& &
-\sum_{A=1}^k\left(Y^i(\phi^{(1)}(t))\frac{\partial^2L}{\partial
q^i\partial v^k_A}\Big\vert_{\phi^{(1)}(t)}+
Y^i_B(\phi^{(1)}(t))\frac{\partial^2L}{\partial v^i_B\partial
v^k_A}\Big\vert_{\phi^{(1)}(t)}\right)
\frac{\partial\phi^k}{\partial t^A}\Big\vert_{t}
\\&=&
Y^i(\phi^{(1)}(t))\frac{\partial L}{\partial q^i}\Big\vert_{\phi^{(1)}(t)} -
Y^i(\phi^{(1)}(t))\frac{\partial L}{\partial q^i}\Big\vert_{\phi^{(1)}(t)}=0 \ ,
\quad \makebox{\rm (on $U_p$)}\ .
\eeann
\qed

\begin{corol}
\label{noethernat}
 If $Z^C\in\vf(T^1_kQ)$ is an infinitesimal natural Cartan symmetry
 of a  $k$-symplectic Lagrangian system $(T^1_kQ,\omega_L^A,E_L)$
  then the functions  $f^A=Z^{V_A}(L)-\zeta^A$, $1\leq A\leq k$,
 define a conservation law  on $U_p$.
\end{corol}
\proof
 In this case,
we have
$$
\inn(Z^C)\theta_L^A=\theta_L^A(Z^C)=\d L\circ
S^A(Z^C)=\d L(Z^{V_A})=Z^{V_A}(L)\;,
$$
 and thus the functions
$f^A$ of Proposition \ref{condnoeth} can be written
$$
f^A=Z^{V_A}(L)-\zeta^A\;, \; 1\leq A\leq k\;.
$$
\qed

The case $k=1$ corresponds to Classical Mechanics, and  the above results can be found in \cite{cra}.

\begin{quote}{\bf Remark}:
 The above Noether's theorem can be rewritten introducing
the following generalization of the so-called {\sl Tulczyjew
operator}  \cite{Tulczy1} for our case:  Let  $g=(g^1, \ldots , g^k):Q\to \Real^k$
be a function, we define $\d_{T}g: T^1_k Q \to \Real$ by
$$
 \d_{T}g({v_1}_q, \ldots, {v_k}_q)=\sum_{A=1}^k {v_A}_q(g^A)=
   v^i_A\,\derpar{g^A}{q^i}
 \quad .
 $$
 Then it is not difficult to prove that the condition
$Z^C(L)= \d_{T}g$ is equivalent to the conditions
$\Lie(Z^C)\theta_L^A=\d\tau^*g^A$ and $Z^C(E_L)=0$. Therefore, by
comparing with item 2 in Proposition \ref{condnoeth} we observe that
the functions $f^A$ can be written as
$$
f^A= Z^{V_A}(L)-\tau^*g^A \ , \quad A=1,\ldots,k\ .
$$
\end{quote}
 Therefore, we have the following proposition,

\begin{prop}
\label{r1}
If $Z \in \vf(Q)$ and $Z^C (L)= d_Tg$,
 where $g=(g^1, \ldots ,g^k)\colon Q\to \Real^k $, then the functions
$f^A= Z^{V_A}(L) -\tau^*g^A$, $1\leq A\leq k$,
 define a conservation law.
\end{prop}
\proof
 This result is a consequence of Theorem \ref{0Nthsec}.
In fact, $ Z^C (L)= d_Tg $ is equivalent to
$\Lie(Z^C)\theta_L^A=\d\tau^*g^A$, and $Z^C(E_L)=0$, which implies
$$
\Lie(Z^C)\omega_L^A=0 \quad \makebox{and} \quad  \Lie(Z^C)E_L=0\;,
$$
that is, $Z^C$ is an infinitesimal natural Cartan symmetry. Then
by Theorem \ref{0Nthsec}, $f=( f^1,\ldots , f^k)$ is a
conservation law.
\qed

In the case $k=1$, this statement can be found  in \cite{Caloma} and  \cite{mssv}.

Finally, we also have that:

 \begin{teor}\label{NthL}
 {\rm (Noether):}
 If $Y\in\vf(T^1_kQ)$ is an infinitesimal Cartan symmetry of a
 $k$-symplectic Lagrangian system $(T^1_kQ,\omega_L^A,E_L)$ then, for every
${\bf\Gamma}=(\Gamma_1,\ldots ,\Gamma_k)\in\vf^k_L(T^1_kQ)$, we have
 $$
\Lie(\Gamma_A)f^A=0 \qquad \mbox{\rm (on $U_p$)}\;.
 $$  \end{teor}
 \proof
This is the same as for Theorem \ref{Nth}.
\qed

\subsection{Equivalent Lagrangians}
  \protect\label{el}

Given a $k$-symplectic Lagrangian system $(T^1_kQ,\omega_L^A,E_L)$,
we know that canonical lifting of diffeo\-morphisms and vector
fields preserve the canonical structures of $T^1_kQ$. Nevertheless,
the $k$-symplectic structure given by the forms $\omega_L^A$ is not
canonical, since it depends on the choice of the Lagrangian function
$L$, and then it is not invariant by these canonical liftings. Thus,
given a diffeomorphism $\Phi\colon T^1_kQ\to T^1_kQ$ or a vector
field $Y\in\vf (T^1_kQ)$, a sufficient condition to assure the
conditions (a) and (b) in definition \ref{nls} would be to
demand that $\Phi$ or $Y$  leave the canonical endomorphisms $S^A$
and the Liouville vector field $\Delta$ invariant (for instance,
$\Phi$ and $Y$ being the canonical lifting of a diffeomorphism and a
vector field in $Q$), and that the Lagrangian function $L$ be also
invariant. In this way, $\omega_L^A$, $E_L$ and hence the
Euler-Lagrange equations are invariant by $\Phi$ or $Y$. However, to
demand the invariance of $L$ is a strong condition, since there are
Lagrangian functions that, being different,  give rise to the same
$k$-symplectic structure $\omega_L^A$, $A=1,\ldots,k$, and the same
Euler-Lagrange equations. Thus, following the same terminology as
in mechanics (see \cite{am}), we can define:

\begin{definition}
\label{gaugeq}
Two Lagrangian functions $L_1,L_2\in\Cinfty (T^1_kQ)$ are
 {\rm gauge equivalent} if
\ben
\item
$\omega_{L_1}^A=\omega_{L_2}^A$, for $A=1,\ldots,k$.
\item
$\vf^k_{L_1}(T^1_kQ)=\vf^k_{L_2}(T^1_kQ)$.
\een
\end{definition}

Gauge equivalent Lagrangians can be also characterized as follows:

\begin{prop}
\label{gaugecarac}
 Two Lagrangians $L_1,L_2\in\Cinfty (T^1_kQ)$  are gauge equivalent if, and only if,
\ben
\item
$\omega_{L_1}^A=\omega_{L_2}^A$, for $A=1,\ldots,k$.
\item
$E_{L_1}= E_{L_2}$,  (up to a constant).
\een
\end{prop}
\proof
 We will prove that, if $\omega_{L_1}^A=\omega_{L_2}^A$, for
$A=1,\ldots,k$, then $\vf^k_{L_2}(T^1_kQ)=\vf^k_{L_1}(T^1_kQ)$ is
equivalent to $E_{L_1}= E_{L_2}$ (up to a constant).

If ${\bf X}=(X_1,\dots,X_k)\in\vf^k_{L_2}(T^1_kQ)=\vf^k_{L_1}(T^1_kQ)$,
then
$$
0= \sum_{A=1}^k\inn(X_A)\omega_{L_1}^A-\d E_{L_1}=
\sum_{A=1}^k\inn(X_A)\omega_{L_2}^A-\d E_{L_2}
$$
but as $\omega_{L_1}^A=\omega_{L_2}^A$, this implies that $\d
E_{L_1}= \d E_{L_2}$, and hence $E_{L_1}= E_{L_2}$, up to a
constant.

Conversely, if
$\omega_{L_1}^A=\omega_{L_2}^A$, and $E_{L_1}=E_{L_2}$ (up to a constant),
then for every ${\bf X}=(X_1,\dots,X_k)\in\vf^k_{L_1}(T^1_kQ)$, we have
$$
0= \sum_{A=1}^k\inn(X_A)\omega_{L_1}^A-\d E_{L_1}=
\sum_{A=1}^k\inn(X_A)\omega_{L_2}^A-\d E_{L_2}
$$
so ${\bf X}\in\vf^k_{L_2}(T^1_kQ)$, and in the same way we prove
that if ${\bf X}\in\vf^k_{L_2}(T^1_kQ)$, then ${\bf X}\in\vf^k_{L_1}(T^1_kQ)$.
 \qed

For gauge-equivalent Lagrangians, definition \ref{gaugeq} guarantees
the invariance of the set of $k$-vector fields which are solution to the
geometric Euler-Lagrange equations (\ref{genericEL}).
Nevertheless,  this condition is also sufficient to assure the invariance
of the set solutions to the Euler-Lagrange equations  (\ref{ELe}).
In fact:

\begin{prop}
If the Lagrangian functions $L_1,L_2\in\Cinfty (T^1_kQ)$ are gauge
equivalent then, the Euler-Lagrange equations (\ref{ELe})
associated to $L_1$ and $L_2$ have the same solutions.
\end{prop}
\proof
 If $L_1,L_2\in\Cinfty (T^1_kQ)$ are gauge equivalent, then
by the Proposition \ref{gaugecarac} we have:
$\omega_{L_1}^A=\omega_{L_2}^A$, for $A=1,\ldots,k$ and $E_{L_1}=
E_{L_2}$,  (up to a constant).
As $\omega_{L_1}^A=\omega_{L_2}^A$, for $A=1,\ldots,k$,  from
(\ref{omegala}) we deduce that
\begin{equation}
\label{gaugeloc}
  \frac{\partial^2 L_1}{\partial q^j\partial v^i_A}=\frac{\partial^2 L_2}{\partial q^j\partial v^i_A}
 \qquad \mbox{\rm and} \qquad
\frac{\partial^2 L_1}{\partial v^j_B\partial v^i_A}= \ds\frac{\partial^2 L_2}{\partial v^j_B\partial v^i_A}\; .
\end{equation}
Therefore, we obtain \bea \label{term1} &&\frac{\partial}{\partial
t^A}\left(\frac{\partial L_1}{\partial
v^i_A}\Big\vert_{\phi^{(1)}(t)}\right)
 \frac{\partial^2L_1}{\partial q^j\partial v^i_A}\Big\vert_{\phi^{(1)}(t)}
\frac{\partial\phi^j}{\partial t^A}\Big\vert_t+
\frac{\partial^2 L_1}{\partial v^j_B\partial v^i_A}\Big\vert_{\phi^{(1)}(t)}
\frac{\partial^2\phi^j}{\partial t^A\partial t^B}\Big\vert_t
\nonumber \\ &=&
\frac{\partial^2 L_2}{\partial q^j\partial v^i_A}\Big\vert_{\phi^{(1)}(t)}
\frac{\partial\phi^j}{\partial t^A}\Big\vert_t+
\frac{\partial^2 L_2}{\partial v^j_B\partial v^i_A}\Big\vert_{\phi^{(1)}(t)}
\frac{\partial^2\phi^j}{\partial t^A\partial t^B}\Big\vert_t=
\frac{\partial}{\partial t^A}\left(\frac{\partial L_2}{\partial v^i_A}\Big\vert_{\phi^{(1)}(t)}\right)\ .
\eea
Furthermore, $E_{L_1}=E_{L_2}$ (up to a constant), then
$\ds\frac{\partial E_{L_1}}{\partial q^j}=\ds\frac{\partial E_{L_2}}{\partial q^j}$,
 and from (\ref{energyL}) we deduce
\begin{equation}
\label{ener}
v^i_A\frac{\partial^2 L_1}{\partial q^j\partial v^i_A}-\frac{\partial L_1}{\partial q^j}=
v^i_A\frac{\partial^2 L_2}{\partial q^j\partial v^i_A}-\frac{\partial L_2}{\partial q^j}\ .
\end{equation}
 From (\ref{gaugeloc}) and (\ref{ener}) we obtain
\begin{equation}
\label{term2}
\frac{\partial L_1}{\partial q^j}=\frac{\partial L_2}{\partial q^j}\ ,
\end{equation}
and then, from (\ref{term1}) and (\ref{term2}) we obtain
$$
\sum_{A=1}^k\frac{\partial}{\partial t^A}\left(\frac{\partial
L_1}{\partial v^i_A}\Big\vert_{\phi^{(1)}(t)}\right)
 -\frac{\partial L_1}{\partial q^j}\Big\vert_{\phi^{(1)}(t)}=\sum_{A=1}^k
\frac{\partial}{\partial t^A}\left(\ds\frac{\partial L_2}{\partial
v^i_A}\Big\vert_{\phi^{(1)}(t)}\right)
 -\frac{\partial L_2}{\partial q^j}\Big\vert_{\phi^{(1)}(t)}\ ,
$$
which implies that $\phi\colon\Real^k\to Q$ is a solution to the
Euler-Lagrange equations associated to $L_1$ if, and only if,
it is a solution to the Euler-Lagrange equations associated with $L_2$.
   \qed

As a generalization of an analogous result in mechanics
(see \cite{am}, p 216), we have the following results:

\begin{prop}
\label{omega=0}
A Lagrangian $L\colon T^1_kQ\to \Real$ satisfies $\omega_L^A=0$, for every
$A=1,\ldots, k$, if, and only if, there exist
 $\alpha^1,\ldots, \alpha^k\in\df^1(Q)$, closed $1$-forms on $Q$ and a  function
$f\in \mathcal{C}^\infty(Q)$, such that $L=\widehat{\alpha}+\tau^*f$
(up to a constant), where $\hat\alpha\in\Cinfty (T^1_kQ)$ is the function defined by
$$
\begin{array}{ccccc}
\hat\alpha&\colon&T^1_kQ&\longrightarrow&\Real
\\ & & w_q=(v_{1_q},\ldots, v_{k_q}) & \mapsto &
\ds\sum_{A=1}^k\alpha^A_q(v_{A_q})
\end{array}\ .
$$
\end{prop}
\proof
  Suppose that $\omega_L^A=-\d\theta_L^A=0\,,\;1\leq
A\leq k$, then $\theta_L^A=\d L\circ S^A$ are closed and semi-basic
$1$-forms on $T^1_kQ$, then $\d L\circ S^A$ are basic forms and
 there exist $\alpha^A\in\df^1(Q)$ such that
\begin{equation}
\label{stau}
 \d L\circ S^A=\tau^*\alpha^A\quad, \quad 1\leq A\leq k \ .
\end{equation}
 Moreover, since $0=\d\theta_L^A=\d(\tau^*\alpha^A)=\tau^*(d\alpha^A)$,
 then $d\alpha^A=0$; that is, each $\alpha^A$ is a closed $1$-form on  $Q$.
Furthermore, by a computation in local coordinates we obtain
 $\d\hat\alpha\circ S^A=\tau^*\alpha^A$, and from (\ref{stau}) we have
 $\d\hat\alpha\circ S^A=\tau^*\alpha = dL\circ S^A$.
 Then $\d(L-\hat\alpha)\circ S^A=0$. Therefore, the $1$-form $\d(L-\hat\alpha)$ is closed and
 semi-basic. As a consequence, $\d(L-\hat\alpha)$ is a basic
 $1$-form; that is, there exist $f\in\Cinfty(Q)$ such that
 $\d(L-\hat\alpha)=\tau^*\d f=\d(\tau^*f)$. Then
 $L=\hat\alpha+\tau^*f$ (up to a constant).

Conversely, let us suppose that $L=\hat\alpha+\tau^*f$ (up to a
constant).  For every $A=1,\ldots, k$ we have
$$
 \theta_L^A=dL\circ S^A=d(\hat\alpha+\tau^*f)\circ S^A=
\d\hat\alpha\circ S^A=\tau^*\alpha^A\;,
$$
 since $\d\tau^*f$ vanishes on the vertical vector fields.
As $\alpha^A$ is closed, $\d\alpha^A=0$ and we obtain
$$
\omega_L^A=-\d\theta_L^A=-\d(\tau^*\alpha^A)=-\tau^*(\d\alpha^A)=0\;.
$$
\qed

\begin{prop}
\label{gaugeL}
The Lagrangian functions $L_1,L_2\in\Cinfty (T^1_kQ)$ are gauge
equivalent if, and only if, $L_1=L_2+\hat\alpha$  (up to a constant).
\end{prop}
\proof
 Let us suppose that $L_1,L_2\in\Cinfty (T^1_kQ)$ are gauge equivalent.
As $\omega_{L_1}^A=\omega_{L_2}^A$, then
$\omega_{L_1-L_2}^A=0$, $1\leq A\leq k$. Thus, by Proposition
\ref{omega=0}, there exist $\alpha^1,\ldots,\alpha^k\in Z^1(Q)$
and $f\in \mathcal{C}^\infty(Q)$ such that
$L_1-L_2=\hat\alpha+\tau^*f$ (up to a constant).

 From Proposition \ref{gaugecarac} we know that $E_{L_1}= E_{L_2}$,  (up
to a constant), or equivalently, $E_{L_1}- E_{L_2}=0$ (up to a constant). Therefore,
$$
\begin{array}{lcl}
0&=&E_{L_1}- E_{L_2}=\Delta(L_1)-L_1
-\Delta(L_2)+L_2=\Delta(L_1-L_2)-(L_1-L_2)\\\noalign{\medskip}&=&
\Delta(\hat\alpha+\tau^*f)-(L_1-L_2)=
\hat\alpha-(L_1-L_2)\quad \makebox{(up to a constant).}
\end{array}
$$

Conversely, let us suppose $L_1=L_2+\hat\alpha$  (up to a  constant).
First, a simple computation gives
 $$
\begin{array}{lcl}
 \omega_{L_2}^A-\omega_{L_1}^A&=&
\d(\theta_{L_1}^A-\theta_{L_2}^A)=d(d(L_1-L_2)\circ S^A)=
\d(\d\hat\alpha\circ S^A)=\d(\tau^*\alpha^A)
\\\noalign{\medskip}&=&\tau^*(\d\alpha^A)=0\,.
 \end{array}
$$
Thus $ \omega_{L_1}^A=\omega_{L_2}^A$.
Furthermore,
 $$
 E_{L_1}=\Delta(L_1)-L_1= \Delta(L_2+\hat\alpha)-(L_2+\hat\alpha)=
  E_{L_2} + \hat\alpha - \hat\alpha= E_{L_2}\quad
\makebox{\rm (up to a constant)},
 $$
 since $\Delta(\hat\alpha)=\hat\alpha$.
As $ \omega_{L_1}^A=\omega_{L_2}^A$ and $ E_{L_1}=E_{L_2}$ (up
to a constant), which means that $L_1$ and $L_2$ are gauge equivalents
 (see Proposition \ref{gaugecarac}).
\qed

\subsection{Lagrangian gauge symmetries}
 \protect\label{lsnt}

Bearing in mind the discussion made in the last section, we can define:

\begin{definition}
Let $(T^1_kQ,\omega_L^A,E_L)$ be a $k$-symplectic Lagrangian system.
\ben
\item
A {\rm Lagrangian gauge symmetry} is a diffeomorphism $\Phi\colon
T^1_kQ\to T^1_kQ$ such that $L$ and $\Phi^*L$ are gauge-equivalent
Lagrangians; that is, $\Phi^*L=L+\hat\alpha$ (up to a constant),
$\hat\alpha\in\Cinfty (T^1_kQ)$ being the function defined in
Proposition \ref{omega=0}.

In the particular case where $\Phi^*L=L$  (up to a constant),
then $\Phi$ is said to be a {\rm Lagrangian strict symmetry}.

A Lagrangian gauge symmetry is said to be {\rm natural} if there exists
a diffeomorphism $\varphi\colon Q\to Q$ such that
$\Phi=(T^1_k)\varphi$.
\item
An {\rm infinitesimal Lagrangian gauge symmetry} is a vector field
$Y\in\vf (\Tan Q)$ whose local flows are Lagrangian gauge symmetries.

In the particular case where $\Lie(Y)L=0$,
then $Y$ is said to be an {\rm infinitesimal Lagrangian strict symmetry}.

An infinitesimal Lagrangian gauge symmetry is said to be {\rm natural} if
there exists a vector field $Z\in\vf (Q)$ such that $Y=Z^C$,
\een
\end{definition}

\begin{quote}{\bf Remark}:
A Lagrangian gauge symmetry $\Phi\colon T^1_kQ\to T^1_kQ$
of a $k$-symplectic Lagrangian system
is not necessarily a Cartan symmetry, since in general
$\Phi^*\omega_L^A\not=\omega_{\Phi^*L}^A$,
for $A=1,\ldots, k$, and
$\Phi^* E_L\not= E_{\Phi^*L}$, as can be easily proved with a simple calculation in coordinates.
\end{quote}
 In general we have:

\begin{lem}
\label{lema3}
 Let $\varphi\colon Q\to Q$ be a diffeomorphism and let
$\Phi=T^1_k(\varphi)$ the canonical prolongation of $\varphi$. Then:
$$
(i)\; \Phi^*\theta_L^A=\theta_{\Phi^*L}^A, \quad(ii)\;
\Phi^*\omega_L^A=\omega_{\Phi^*L}^A,\quad (iii)\;
\Phi^*E_L=E_{\Phi^*L}\;.
$$
\end{lem}
\proof
 This is a direct consequence of Lemma \ref{lema.2} and the
definition of  $\theta_L^A$. In fact, for $\Phi=T^1_k(\varphi)$ we obtain
 \beann
 \Phi^*\theta_L^A&=& \Phi^*(\d L\circ S_A)=\d (\Phi^*L)\circ
 S_A)=\theta_{\Phi^*L}^A\;.
\\
\Phi^*\omega_L^A&=&\Phi^*(-\d\theta_L^A)=-\d\Phi^*\theta_L^A=\omega_{\Phi^*L}^A\;.
 \\
\Phi^*E_L&=&\Phi^*(\sum_{A=1}^k\Delta_A(L)-L)=\sum_{A=1}^k\Delta_A(\Phi^*L)-\Phi^*L=E_{\Phi^*L}\;.
\eeann
 \qed

And then we have the following relation between natural Cartan symmetries
and natural gauge symmetries:

\begin{prop}
Let $(T^1_kQ,\omega_L^A,E_L)$ be a $k$-symplectic Lagrangian system.
Then, $\Phi\colon T^1_kQ\to T^1_kQ$ is a natural Cartan symmetry
if, and only if, it is a natural Lagrangian gauge symmetry.
\end{prop}
\proof
 If $\Phi=T^1_k(\varphi)$ for some diffeomorphism $\varphi\colon Q\to Q$,
by lemma (\ref{lema3}) we have that
$$
\Phi^*\omega_L^A=\omega_{\Phi^*L}^A \quad , \quad
\Phi^*E_L=E_{\Phi^*L}
$$
therefore
$$
\left. \begin{array}{ccc}
\Phi^*\omega_L^A&=&\omega_L^A \\
\Phi^*E_L&=&E_L
\end{array}\right\}
\ \Longleftrightarrow \
\left\{ \begin{array}{ccc}
(\omega_{\Phi^*L})_A&=&\omega_L^A \\
E_{\Phi^*L}&=&E_L
\end{array}\right.
$$
that is, $\Phi$ is a natural Cartan Lagrangian symmetry if, and only if,
 $L$ and $\Phi^*L$ are gauge equivalent Lagrangians and
thus $\Phi$ is a natural Lagrangian gauge symmetry.
  \qed

This result also holds for infinitesimal Lagrangian symmetries,
taking the corresponding local flows.

Finally, we can state a particular version of Noether's theorem for
natural Lagrangian strict symmetries:

\begin{teor}
 {\rm (Lagrangian Noether):}
 If $Y\in\vf((T^1_k)Q)$ is an infinitesimal natural Lagrangian strict symmetry of a
 $k$-symplectic Lagrangian system $(T^1_kQ,\omega_L^A,E_L)$,
with $Y=Z^C$, for some $Z\in\vf(T^1_kQ)$, then the functions
$f^A=Z^{V_A}(L)$, for $1\leq A\leq k$, define a conservation law
$f=(f^1,\ldots,f^k)$.
 \label{NthLs}
\end{teor}
\proof
This is a straightforward consequence of the above proposition
and corollary \ref{noethernat} since in this case,
$$
\d\zeta^A=\Lie(Y)\theta_L^A=\Lie(Z^C)\theta_L^A=0\quad ,\quad 1\leq A\leq k\ .
$$
\qed

In the case $k=1$, the above result can be found in \cite{arn,mssv}.

\section{Conclusions and outlook}

We analyze several kinds of symmetries that can be defined for
Hamiltonian and Lagrangian first-order classical field theories, in their
$k$-symplectic formulation.

First, we define the concept of symmetry (and infinitesimal
symmetry). Second, according to Olver, we define conservation
laws and investigate the problem of associating
conservation laws with symmetries. In this way we have considered
Cartan symmetries (which preserve the $k$-symplectic structures and
physics; i.e., the Hamiltonian or the energy function) and, in
particular, those called ``natural'', which are canonical
liftings of diffeomorphisms or vector fields. We prove that
Cartan symmetries are symmetries and that there is a natural way of
associating them with conservation laws by means of Noether's
theorem. We state and prove this theorem  in different
situations for the Hamiltonian and Lagrangian cases.

Finally, we study and characterize gauge equivalent Lagrangians,
leading to the introduction of Lagrangian gauge symmetries
(which transform a Lagrangian into another equivalent one),
proving that natural Lagrangian gauge symmetries are the same
as natural Cartan symmetries, and stating the corresponding
Noether's theorem.

Further research will be devoted to extending all  these concepts and
results to the $k$-cosymplectic formalism of  first-order classical field theories.

\subsection*{Acknowledgments}

We acknowledge the partial financial support of the project
MTM2006-27467-E/.
The author NRR also acknowledges the financial support of
{\sl Ministerio de Educaci\'on y Ciencia}, Project MTM2005-04947.
We thank Mr. Jeff Palmer for his assistance in preparing the
English version of the manuscript.

\end{document}